# Methodological variations in lagged regression for detecting physiologic drug effects in EHR data

Matthew E. Levine, BA* [a, b]
David J. Albers, PhD [a, b]
George Hripcsak, MD, MS [a, b, c]

[a] Department of Biomedical Informatics
Columbia University Medical Center
622 W. 168th Street, Presbyterian Building 20th Floor
New York, NY 10032

[b] Observational Health Data Sciences and Informatics (OHDSI), New York, NY

[c] NewYork-Presbyterian Hospital, New York, NY
622 W. 168th Street
New York, NY 10032

*Please address correspondence to mel2193@cumc.columbia.edu

# ABSTRACT

We studied how lagged linear regression can be used to detect the physiologic effects of drugs from data in the electronic health record (EHR). We systematically examined the effect of methodological variations ((i) time series construction, (ii) temporal parameterization, (iii) intra-subject normalization, (iv) differencing (lagged rates of change achieved by taking differences between consecutive measurements), (v) explanatory variables, and (vi) regression models) on performance of lagged linear methods in this context. We generated two gold standards (one knowledge-base derived, one expert-curated) for expected pairwise relationships between 7 drugs and 4 labs, and evaluated how the 64 unique combinations of methodological perturbations reproduce the gold standards. Our 28 cohorts included patients in the Columbia University Medical Center/NewYork-Presbyterian Hospital clinical database, and ranged from 2,820 to 79,514 patients with between 8 and 209 average time points per patient. The most accurate methods achieved AUROC of 0.794 for knowledge-base derived gold standard (95%CI [0.741, 0.847]) and 0.705 for expert-curated gold standard (95% CI [0.629, 0.781]). We observed a mean AUROC of 0.633 (95%CI [0.610, 0.657], expert-curated gold standard) across all methods that re-parameterize time according to sequence and use either a joint autoregressive model *with* time-series differencing or an independent lag model *without* differencing. The complement of this set of methods achieved a mean AUROC close to 0.5, indicating the importance of these choices. We conclude that time-series analysis of EHR data will likely rely on some of the beneficial pre-processing and modeling methodologies identified, and will certainly benefit from continued careful analysis of methodological perturbations. This study found that methodological variations, such as pre-processing and representations, have a large effect on results, exposing the importance of thoroughly evaluating these components when comparing machine-learning methods.

**INTRODUCTION**

Widespread adoption of electronic health records (EHRs) over the past 30 years has created a rich resource of observational health data, and research communities continue to dedicate themselves to leveraging these data to improve clinical care and knowledge [1]. EHR-based observational research enables new discoveries that are nearly impossible to achieve using traditional experimental methods, and encourages collaborative, open science [2]. However, in order to properly leverage EHR data in observational studies, we must address the special properties of EHR data by adapting and re-inventing existing statistical methods. Here we formulate how to use lagged linear vector regression with EHR data, using interactions between medication administration and laboratory measurements as our clinical context.

Using EHR data to tackle the identification and characterization of the physiologic effects of drugs is a substantial challenge. Although most drugs have known mechanisms of intended action, the full diversity of their myriad effects on biological function is poorly understood and impractical to study experimentally. Such an understanding is important in the context of adverse effects, where drugs induce unexpectedly harmful consequences, as well as for uncovering beneficial effects not detected in small, controlled clinical trials.

There exist data-driven solutions for studying drugs and their physiologic effects, but challenges remain for uncovering their true complexity. Traditional epidemiological approaches are most successful for identifying relatively simple trends (e.g. does condition X occur after the first administration of drug Y), and progress has been made in automatically detecting adverse drug effects [3] using structured clinical databases [4], clinical notes [5], [6], and online health forums [7]. Recent work has focused on scaling these methods to massive data sets [8] and incorporating all available drug and outcome data. Yet finer temporal structure is often desired in order to better understand and predict physiologic treatment responses.

Computational methods exist for uncovering detailed temporal relationships between drugs and outcomes in EHR data, and recent advances have been made in machine-learning approaches to phenotyping [9], [10], pattern discovery [11]–[13], temporal abstraction over intervals [14], and dynamic Bayesian networks [15]. However, these advances typically highlight one or two approaches at a time, and do not rigorously justify or study methodological decisions that may be inconsequential or vital to a method's success. In addition, many of these approaches rely on assumptions of stationarity that are frequently broken by clinical data [16], [17], or do not account for health care process effects.

Hripcsak et al. [18], [19] have demonstrated that time-series methods applied to EHR data can identify meaningful, high-fidelity [20] trends that relate drugs and physiologic processes. However, standard time-series analysis tools rely on assumptions like stationarity and, to a lesser extent, regular sampling frequencies, which are generally absent from EHR. We have shown that temporal re-parameterizations can overcome non-stationarity, intra-patient normalization can filter out inter-patient effects, and adding contextual variables can address health care process effects [21]–[24]. We nevertheless lack an understanding of how such specific modeling choices—performed alone or in combination—impact inference quality and predictive performance within a lagged linear paradigm for analyzing EHR data.

We consider six important steps in time-series modeling of EHR data, and apply two specific perturbations to each of the following: (i) time series construction [21], [22], (ii) temporal parameterization [23], (iii) intra-subject normalization [21], [22], (iv) differencing (lagged rates of change achieved by taking differences between consecutive measurements) [24], (v) explanatory variables [24], and (vi) regression models.

Here, we systematically evaluate these methodological perturbations in a combinatorial set of 7 drug and 4 lab conditions, and compute a bootstrapped estimate of predictive performance with respect to gold standard expectations for each of the 28 pair-wise relationships under each of 64 ($2^6$) methodological variations. In this way, we probe for modeling choices that provide statistically meaningful improvements to detecting physiologic drug effects. Furthermore, we obtain a more reliable estimate for the ability of well-tuned lagged linear methods to predict physiologic drug effects.

## MATERIALS AND METHODS

*Cohort Criteria*

The 30-year-old clinical data warehouse at NewYork-Presbyterian Hospital, which contains electronic health records for over 5 million patients, was used to examine pairwise relationships between drug order records and laboratory measurements. We selected 7 medications, amphotericin B, simvastatin, warfarin, spironolactone, ibuprofen, furosemide, allopurinol and 4 blood laboratory measurements, total creatine kinase, creatinine, potassium, hemoglobin, resulting in 28 drug-lab pairwise analyses (descriptions are listed in Supplementary Table 2). For each drug-lab pair we identified a cohort of patients that met the following criteria: 1) at least 2 of the laboratory measurements of interest on record, 2) at least 1 order for the drug of interest, and 3) more than 30 combined data points between laboratory measurements of interest and total drug orders for any drug. We collected the entire drug-order history, the entire history of laboratory measurements of interest, and entire history of inpatient admissions for each included patient (for use as optional contextual variables). These selection criteria returned between 2,820 and 79,514 patients for the 28 cohorts, with between 8 and 209 average time points per patient, and between 78,624 and 6,107,601 total time points overall.

*Building a time series with clinical data*

We convert binary inputs to continuous values as follows. We constructed a time series of drug values by setting all drug-orders of interest to 1, and all orders of other drugs to 0 [22], and constructed a time series for contextual variables (in this case, inpatient admission) by setting the event to 1, and setting a 0 at 24 hours before and after that event [24].

Since measurements were sparse and rarely aligned, we interpolated each time series (see Figure 1 for a graphical depiction). For every time point where there was a concept (lab, drug, or inpatient admission), the values of each other variable at that time point were interpolated as the clock-time weighted mean of the preceding and succeeding value of each respective variable. Weighting our interpolation by clock-time allows an estimated lab value at the time of a drug order to be closest to the nearest lab value, and takes into account the trend of the lab near that time. This was performed at each time-point by weighting the nearest two bordering concept values according to their temporal distance from the interpolated time-point. Ultimately, all concepts, whether from categorical or real-valued sources, took on continuous values that were paired at each time point. For a more complete description of how we construct a multivariate time series from clinical data, see our previous work [22], [24].

*Methodological variations for lagged linear regression with clinical data*

In order to evaluate time series methods for uncovering physiologic drug effects, we focused on lagged linear regression and performed 64 ($2^6$) perturbations of the standard methodology. The data we use are nonstandard, biased by the health care process, non-stationary, irregularly measured, and missing not at random, requiring methodological explorations to understand how to cope with irregularities of EHR data [23], [25]–[29]. We consider temporal parameterization, time series window construction, intra-subject normalization, differencing, inclusion of other variables (e.g. related to health care process), and choices in how regression models are computed.

*Temporal parameterization*
Previous studies have shown that indexing a clinical time series by its sequence order can have significant advantages over traditional clock-time [23]. To test this, we indexed our lagged analysis with respect to both real-time and sequence-time. Clock-time was converted to sequence-time by setting all time intervals between interpolated, pre-processed values to unit 1 length, making all times ordered integers with no missing times. For further details on their implementation, see our previous descriptions [23], [24].

*Binning and windowing*
In signal processing, window functions are often used to extract a smoothed or filtered segment of a time series near a particular time point. They are typically non-negative and smooth over a finite interval; examples include a constant over a rectangle, a triangle, and a Gaussian window. The right choice of window function can remove bias from a signal, and can improve results of cross-correlation analysis. However, choosing appropriate windows is challenging and problem-dependent, and improper choices can lead to spurious signals, aliasing, and other spectral leakage pathologies [30]–[33].

We hypothesize a particular type of bias that we introduce in our timeline construction methodology, and attempt to remove it with a simple window function, a maximum function over a 24hr width on the drug time series, which we refer to as "binning". The heuristics we have used previously [22] cause drug signals to diminish when a drug of interest is consistently ordered between two other drugs. Ideally, the drug timeline should retain mass for as long as a patient is consistently taking a drug. We attempted to remove this bias by setting all drugs within 12 hours of the drug of interest to 1. It should be clear that this process is equivalent to applying a fixed-width window equipped with the max-function.

*Regression Models*
We considered lags from 1-30 days when using real-time, and 1-30 indices when using sequence time. We studied two variations of lagged linear regression—univariate (i.e. lags estimated one at a time, independently) and multivariable (i.e. lags estimated jointly). Independent, univariate estimation provides a simple model similar to lagged correlation that separately relates each lagged time-point of each lagged variable to the target response variable; joint, multivariable estimation is an autoregression (specifically, an ARX model) and computes each lagged coefficient conditional on the other estimates, balancing the shared information across lags and thus bringing out more subtle details of each lag. First, we considered *independent* estimation of lagged drug coefficients, $\beta_\tau$, from the following model, where $y_t$ is the lab value (i.e., the outcome of interest) at time $t$, $x_t$ is the drug value at time $t$, and $\tau$ is the lag time (for $\tau = 1:30$):

$$y_t = c_\tau + \beta_\tau x_{t-\tau} + \varepsilon_\tau$$

Second, we considered *joint autoregressive* estimation of lagged drug coefficients, $\beta_\tau$, by the following form (L=30):

$$y_t = c + \sum_{\tau=1}^{L} \beta_\tau x_{t-\tau} + \sum_{\tau=1}^{L} \alpha_\tau y_{t-\tau} + \varepsilon$$

This form generalizes to an arbitrary number of other lagged explanatory variables, $u^i$ (which can include $y$), as:

$$y_t = c + \sum_{\tau=1}^{L} \beta_\tau x_{t-\tau} + \sum_{i=1}^{N} \sum_{\tau=1}^{L} \omega_\tau^i u_{t-\tau}^i + \varepsilon$$

*Differencing*
In time series analysis, pre-processing steps, like taking differences between consecutive measurements, are often performed to de-correlate lagged variables [34]. More formally, a differencing operator can be applied to resolve non-stationarity that results from a unit root in the characteristic equation of an autoregressive stochastic process—the presence of a unit root can be identified with statistical tests, like Dickey-Fuller [35], and removed by iterative differencing [36]. When unit roots remain, ordinary least squares estimation of autoregression coefficients has been shown to fail [37] and non-stationarity persists. The simplest example is the case of a random walk, in which each position is highly correlated with the previous positions. By taking the differences between consecutive steps of a random walk, these correlations are removed and the statistics of the signal can be more easily recovered. Similar effects can be seen in clinical data, where treatments often drive physiologic change. Levine et al. [24] demonstrated that taking differences is an important step in multivariable lagged regression with clinical data; here, we tested the value of differencing in additional clinical and methodological contexts.

*Intra-patient normalization*
Previous work demonstrated that intra-patient normalization is an important step when extracting correct physiologic drug effects using lagged correlation [22]. In order to investigate the importance of removing inter-patient effects in different methodological contexts, we included the option to normalize each patient's time series by subtracting their mean and dividing by their standard deviation. More sophisticated schemes for approaching this problem exist (e.g. Box Cox transform [38] or other power transforms), but we wished to first examine a simpler method. It is also important to note that the univariate lagged regression coefficient (i.e. AR-1) on normalized (zero mean, unit variance) time series is identical to the coefficient from lagged correlation. Thus, as various pre-processing and analytic steps are combined, the resulting method often devolves into a specially named sub-class of methods.

*Including context variables*

In order to account for health care process effects and biases, we often wish to include potential confounding variables in the model. Levine et al. [24] found that including inpatient admission events as autoregressive variables in a multivariable multi-lag model (i.e. vector autoregression [34], [39]) attenuated some confounded physiologic signals. We evaluated the same approach here, and introduced the context variable $z$ to correct lagged drug coefficients, $\beta_\tau$:

$$y_t = c + \sum_{\tau=1}^{L} \beta_\tau x_{t-\tau} + \sum_{\tau=1}^{L} \alpha_\tau y_{t-\tau} + \sum_{\tau=1}^{L} \gamma_\tau z_{t-\tau} + \varepsilon$$

*Gold Standard Creation*

In order to evaluate computationally determined interactions between each drug-lab pair, we created two gold standards for whether a given drug is expected to increase, decrease, or have no effect on a given lab: 1) a *knowledge-base derived gold standard* that was created by synthesizing existing medical literature and knowledge bases—this represents information that could, in theory, be obtained automatically, and 2) a *clinical expert curated gold standard*, for which the knowledge-base derived gold standard was reviewed and edited by a clinical expert. In table 1, we indicate whether a given drug is expected to increase, decrease, or have no effect on a given lab (denoted as 1, -1, 0, respectively), according to the two gold standards (68% total agreement, Cohen's Kappa=0.53, 95% CI [0.27-0.78]).

*Literature search for expected physiologic drug effects*

For each drug-lab pair, an author (ML) searched PubMed for articles using the drug and lab as keywords, along with terms "increase", "decrease", and "association". The authors selected articles that reported quantitative information about associations and causations between the two entities within their abstracts. The author then read these articles and determined whether their reported associations between the drug and lab of interest should be expected to generalize to a large EHR database (e.g., a study of cancer patients would not be included).

*LAERTES knowledge base queries for expected physiologic drug effects*

The LAERTES (Large-scale Adverse Effects Related to Treatment Evidence Standardization) [40] knowledge-base was developed as part of the Observational Health Data Sciences and Informatics initiative to record existing pharmacosurveillance knowledge that could be compared to new empirical evidence. It draws from package inserts, Food and Drug Administration databases, and also the literature. LAERTES was queried for associations between side effects associated with the 4 lab measurements (muscle weakness and rhabdomyolysis for creatine kinase, renal impairment for creatinine, hyperkalemia and hypokalemia for potassium, and anemia for hemoglobin) and each of the 7 drugs.

*Knowledge-base derived gold standard—combining results from literature search and knowledge base*

Resulting directional associations from LAERTES were taken in union with the directional associations from our literature search. When one search method yielded no associations, and the other did, we took the association, rather than the null result (except in the case of ibuprofen and total creatine kinase, for which we rejected LAERTES's positive result). When multiple results were present in the LAERTES results, we selected those that matched results in the literature—this occurred twice, for spironolactone's effect on potassium and ibuprofen's effect on potassium. Together, these data formed the knowledge-base derived gold standard.

*Expert-curated gold standard*

A clinical expert (GH) subsequently curated the knowledge-base derived gold standard, and modified 9 of its 28 expected associations. The expert modified the directionality only twice (i.e. -1 to +1), where he believed that diuretic-induced anemia was less likely to be present than rises in hemoglobin due to diuretic-induced fluid loss. The other seven modifications removed expected effects in the knowledge-base derived gold standard (i.e. changed +1 or -1 to 0), which the expert judged sufficiently rare to be missing from a database of the size of ours.

### *Evaluating accuracy of lagged regressions*

We evaluated the predictive accuracy of each tested method and the associated uncertainty by performing a layered bootstrap resampling over patient cohorts [41]. Figure 2 provides a schematic for the experimental protocol.

*Estimating variance of lagged drug coefficients*

We empirically computed estimates of variance for the lagged drug coefficients, $\beta_\tau$, using a bootstrap estimate of variance. For each drug-lab cohort, we sampled patients with replacement to create 200 bootstrapped samples, and ran all 64 regressions for each of these 200 samples from the drug-lab cohort. We estimated the variance of $\beta_\tau$ using the variance of these samples, and subsequently determined empirical 95% confidence intervals of $\beta_\tau$ ($[\beta_\tau - 1.96\sigma, \beta_\tau - 1.96\sigma]$, where $\sigma$ is the standard deviation of the samples of $\beta_\tau$).

*Classifying lagged drug coefficient profiles*

We are ultimately interested in the trajectory of $\beta_\tau$ as they vary over $\tau$, and write $\boldsymbol{\beta} = \{\beta_\tau\}_{\tau=1}^{30}$. In order to perform a first-order evaluation of lagged drug coefficients trajectories, we first converted them to the format of the gold standards (increase, decrease, or no effect). We classified $\boldsymbol{\beta}$ as increasing (+1) if at least 15 consecutive coefficients were all greater than zero within 95% confidence interval, decreasing (-1) if at least 15 consecutive coefficients were all less than zero within 95% confidence interval, and neither (0) otherwise. We selected 15 as the threshold because it is half-the number of total estimated coefficients, making it the smallest threshold that can ensure there will be only one directional designation (we did not want a trajectory of $\boldsymbol{\beta}$ to be classified as both increasing and decreasing).

*Computing predictive performance of lagged regressions with respect to gold standards*
For each of the 64 method combinations, we evaluated classifications of the 28 gold standard drug-lab effects by estimating a Receiver Operating Characteristic (ROC) curve, and reported the area under ROC (AUROC) separately for the two gold standards. Recall that AUROC is a common evaluation metric for binary classification models, and is equal to the expected probability that the model will rank a randomly chosen positive event above a randomly chosen negative one.

Given our ranked classifications (-1,0,1), we evaluated sensitivity and specificity of each method's ability to perform binary discrimination across two thresholds, -0.5 and 0.5, which provided two points for an ROC curve. We computed AUROC using simple trapezoidal integration.

*Estimating variance of AUROC for each methodological variation*
In order to estimate the variance of each method's AUROC, we leveraged the previously performed bootstrapped regressions. For each of the 200 previously computed estimates of $\boldsymbol{\beta}$ for each drug-lab pair, we created a new classification using a confidence interval with fixed variance (previously computed) that was centered at that particular bootstrapped estimate of $\boldsymbol{\beta}$.

We thus obtained 200 independent samples of $\boldsymbol{\beta}$. We classified these, and subsequently arrived at 200 independent, identically applied samples of AUROC for each methodological variation, which were used to statistically compare performance of different methods.

*Comparing predictive performance of lagged regression methods*
We want to compare disjoint classes of methods; for example, we want to compare all methods that use sequence time against all methods that use real time, and ask whether sequence time or real time offers an average performance benefit. In order to perform such comparisons, we report the average difference between AUROCs and the 95% Confidence Intervals (CIs) of this difference for both gold standards. Concretely, we compared two disjoint groups of methods by computing the difference between each group's mean AUROC. We then estimated the 95% CI of this difference using the variance of the pairwise differences between each group's 200 mean sampled AUROCs. This results in a determination of whether one disjoint group of methods is better or worse than another, within a 95% CI, and enables queries like "overall, is it better to use sequence time or real time?" or "overall, is it better to use sequence time with or without normalization?".

We performed these comparisons systematically to arrive at a final set of statistically significant methodological variations. First, we evaluated the impact of each variation across all other variations, i.e. *marginal impact*; for example, we compared all methods that use normalization against all methods that do not. Then we evaluated the impact of each variation given a variation of another variable; for example, we compared all methods that use normalization and sequence time against all methods that use normalization and real time. We also compared sequences of 3 variations. This allowed us to evaluate the impact of methods, both alone and in combination. We report methodological variations that are influential alone and in combination with others, along with the magnitude of their marginal impact on AUROC.

*Summary*

In order to evaluate and compare methodological variations of lagged linear regressions for determining physiologic drug effects from clinical time series, we 1) identify patient cohorts for each drug-lab pair of interest, 2) report the predictive performance of each method with respect to two gold standards, and 3) draw statistically meaningful comparisons between classes of methods to demonstrate important modeling steps that ought to be taken either alone or in combination to achieve desired results.

# RESULTS

## *Illustrating example of importance of methodological variations of lagged linear regression for assessing physiologic drug effects*

In order to illustrate the importance of variations in methodology for analysis of clinical time series, we examined some possible inferences of the relationships between amphotericin B and levels of potassium and creatinine. Our knowledge-base derived gold standard and clinical expert agreed that amphotericin B should be expected to raise creatinine levels and lower potassium levels.

Figure 3 shows the resulting inferences when varying three aspects of the computation (temporal parameterization, differencing, and regression models) and fixing the other three aspects (normalization, no additional context variable, and no binning). Figure 3a shows that the expected trends are accurately reconstructed with statistical significance when using sequence time, differences, and a joint AR model. Figures 3b and 3c show that no significant association can be found when switching to real-time or not using differences. However, Fig 3d shows that multiple changes to the successful method in Fig 3a (using an independent lag model *and* not using differences) can obtain expected trends, albeit with less significance for creatinine (blue). Results from these methodological combinations for all 28 drug-lab pairs are shown in Supplementary Figures 1-7.

## *Combinatorial evaluation of lagged regression assessments of physiologic drug effects under methodological variations*

In order to thoroughly understand the impact of methodological choices in this context, we evaluated all 64 combinations of methods with respect to the two gold standards (knowledge-base derived, and expert-curated).

Our main results are shown in Figure 4. We report each method's AUROC and an estimate of the AUROC variance for both gold standards; we rank the results by descending expert-curated AUROC, and indicate the vector of method pairings for each row in the plot. These results are also enumerated explicitly in Supplementary Table 1.

We first point out that, surprisingly, the majority of method combinations had AUROC of 0.5, indicating performance no better than chance. This implies that the choice of methods, combinations of methods, and even the data representation—differences versus raw values—is very important. Furthermore, while the two gold standards differed significantly according to Table 1, they agreed fairly well on which combinations were better than chance. The superior performance of some combinations does not appear to be artifact.

We observe that there is a concentration of methods using sequence time at the top of the plot, suggesting that sequence time is a beneficial choice independent of other methods. We can also observe patterns that relate differencing with model choice—in particular, we note that of our four possible combinations of differencing and model, only two of these (differences with joint estimation and no differences with independent estimation) ever yield AUROC above 0.5. This suggests an interaction between these two choices, which we subsequently interrogate quantitatively.

The best method, according to the expert-curated gold standard used sequence time, normalization, differencing, a joint AR model, no binning, and no additional context variable (AUROC = 0.705, 95%CI [0.629, 0.781]). According to the knowledge-base derived gold standard, the best method also used sequence time, normalization, and no additional context variable, but did not use differencing, used an independent lag model, and used binning (AUROC = 0.794, 95%CI [0.741, 0.847]).

*Comparing predictive performance between lagged regression methods*
We test for statistically significant differences between marginal effects of different method variations. We observe that choosing sequence time instead of real time is the only single methodological choice that both gold standards agree has a statistically significant effect. For the knowledge-base derived gold standard, sequence time yields a 0.049 (95%CI [0.035, 0.063]) marginal AUROC improvement over real time; for the expert-curated gold standard, the marginal improvement is 0.050 (95%CI [0.038, 0.062]).

In addition, we examined combinations of method choices, and found a consistent, statistically significant indication that a joint AR model is *better* with differences than without (0.062 marginal AUROC improvement with 95%CI [0.045, 0.079] for knowledge-base derived gold standard, 0.074 marginal AUROC improvement with 95%CI [0.053, 0.094] for expert-curated gold standard), and that an independent lagged model is *worse* with differences than without (0.083 marginal AUROC reduction with 95%CI [-0.100, -0.065] for knowledge-base derived gold standard, 0.094 marginal AUROC reduction with 95%CI [-0.114, -0.075] for expert-curated gold standard). We also evaluated the converse statements (e.g. when using differences, is joint AR or independent lag model significantly better), and found similar associations.

We further compared the two preferred pairs, and found that while the independent lag model without differences slightly outperformed the joint AR model with differences overall (0.021 marginal AUROC improvement for both gold standards), these changes were not statistically significant (95%CI [-0.046,0.004] for knowledge-base derived gold standard, 95% CI [-0.049,0.008] for expert-curated gold standard). However, the opposite, albeit statistically insignificant, effect was observed when comparing these methods only in the context of the clearly preferred sequence time.

We ultimately found that once a choice of sequence time and either of the preferred pairs of differencing and modeling (i.e. no differences with independent lag model or differences with joint AR model) was made, no additional choices (binning, context variables, normalization) provided marginal improvement to AUROC with statistical significance. We observe a mean AUROC of 0.633 (95%CI [0.610, 0.657]) for expert-curated gold standard (and 0.622 mean AUROC with 95%CI [0.603, 0.641] for knowledge-base derived gold standard) across methods that use sequence time and one of the preferred difference-model pairs, whereas the complement of this set of methods achieves a mean AUROC close to 0.5 (0.512 with 95%CI [0.506, 0.517] for clinically curated-gold standard; 0.507 with 95%CI [0.503, 0.512] for knowledge-base derived gold standard). In this way, we demonstrate that temporal parameterization, time series differencing, and regression-type are important choices that must be selected in concert to achieve optimal predictive performance.

*Comparing evaluations from two gold standards*
The gold standards differed on 32% of cases (Cohen's Unweighted Kappa=0.53, 95% CI [0.27-0.78]; Cohen's Linear Weighted (ordinal) Kappa=0.54, 95% CI [0.11-0.97]). In two cases, the knowledge-base derived gold standard reported diuretics as possibly causing anemia, thus lowering hemoglobin, without accounting for potential diuretic fluid loss and resultant rise in hemoglobin. This represents a disconnect between the condition (anemia) and the observed entity (hemoglobin), which was noted by the expert. In other cases, a potential side effect was judged to be sufficiently rare that it should be missing from a database of the size of ours.

The effect of the difference in gold standards can be seen in Figure 4 and Figure 5. Figure 5 shows that the AUROCs for each methodological variation are correlated across the two gold standards (Pearson correlation coefficient 0.759, 95%CI [0.631, 0.847]), but that substantial differences exist. Figure 4 shows that each gold standard would rank individual methods differently; nevertheless, major conclusions of the study, such as the superiority of using sequence time and the dependencies between differencing and regression-type, are upheld by both gold standards.

**DISCUSSION**

Here we study how lagged linear regressions, a simple, robust, commonly used class of methods, can be tuned to efficiently extract drugs' temporal effects on patient physiology from EHR data. Data in the EHR present a variety of challenges (low, erratic measurement frequency, high noise, and non-stationarity), making time-series analysis highly non-trivial and requiring careful pre-processing and re-parameterization. We evaluated combinations of pre-processing, modeling, and temporal parameterization steps in order to understand how to better cope with challenges in extracting temporal information from EHR data. We used 64 of these methodological perturbations to analyze 28 drug-lab pairs, and evaluated the results against two gold standards.

We found that the correct combination of regression type (independent lag or joint autoregressive) and differencing was essential—independent lag models cannot be used with differencing, whereas the joint AR model must be used differencing. Furthermore, we found a large significant improvement (for expert-curated gold standard, 0.05 average AUROC increase, 95%CI [0.038, 0.062]) when re-indexing time according to the sequence of events. These selections created high-performing methods, and the top methods achieved AUROC of over 0.7 (for knowledge-base derived gold standard, best AUROC = 0.794, 95%CI [0.741, 0.847]; for expert-curated gold standard, best AUROC = 0.705, 95% CI [0.629, 0.781]).

We also found that the regressions were robust to our choices of normalization, binning, and context variable inclusions. While these choices were statistically unimportant, in aggregate, among our cohorts, their impact could become more noticeable when testing different hypotheses or when using different data. Moreover, we selected one simple form for each of these variations, and it is likely that more targeted formulations will have greater effects.

*Benefits of multiple gold standards*
Gold standards often vary, but by using several gold standards, researchers can—formally or informally—assess their evaluations' sensitivities to the gold standard. If only one gold standard is used, then there is no way to characterize the dependency of conclusions on that particular gold standard. In our case, results were similar but not identical for the two gold standards, indicating that our findings are not mere artifacts of the gold standard. It is important to note that our gold standards were not completely independent, as one author created the knowledge-base derived gold standard, and the clinical expert modified it according to clinical and informatics knowledge.

*Reflections on important methodological steps*
We found that decomposing the overall modeling process into smaller, discrete methods allowed us to systematically interrogate the effect of each choice. However, it is also instructive to note that many of the combinations of methods are in fact equivalent to established methodologies. For example, the joint autoregressive model is very similar to Granger causality, and the independent lag model is analogous to lagged correlation analysis up to normalization. Both of these modeling methods, combined with any windowing function, fall under similar classes of statistical spectral analysis methods and econometrics [34], [36], [42]–[44].

Our previous studies have reported improved performance of lagged methods on EHR data when using sequence time [23], [24], and have investigated the mechanics of these phenomena [16], [17], [23]. We maintain the hypothesis that sequence time removes non-stationarity by leveraging the fact that clinicians sample at rates proportional to patient variability [23], [45], but feel that this hypothesis, while implied, has yet to have been explicitly proven. Lagged regression methods rely on assumptions of weak stationarity, and their performance improves when data are pre-processed to remove temporal swings in mean and variance. There exist methods like autoregressive moving average models that can cope with certain relatively benign non-stationarity effects, such as a slowly and continuously varying mean, but these models are likely unable to resolve clinical non-stationarity effects that are combined with data missing non-randomly (e.g., correlated with health). Such EHR-data-specific pathologies were the original motivation for even attempting sequence time-based methods.

Non-stationarity in EHR data may partly manifest in unit roots of the characteristic equation of the autoregressive stochastic process, causing failure of ordinary least squares estimation, and ought to be explicitly tested in the future using the augmented Dickey-Fuller test [34], [35]. While we optionally applied a differencing operator once to our clinical time series, we did not test for the presence of unit roots. Future work may benefit from iteratively applying a difference operator and re-testing with a statistical test, like augmented Dickey-Fuller, until unit roots are removed, as is the strategy of the Box-Jenkins modeling approach [36].

Differencing is a well-known method [34] for reducing correlation between lagged variables in time-series analysis, and Levine et al. [24] provided anecdotal evidence of its benefit for lagged linear analysis of drug and lab data from the EHR. For this reason, we expected it to improve results across all methods. We were surprised to learn that differencing corrupted the performance of the independent lag model. We recognize, however, that there is a tradeoff between sharing uncorrelated information across variables and adding noise to any particular variable. In the case of the independent lag model, we correlate with one variable at a time, effectively losing all of the upside of differencing. Because the joint autoregressive model holds some advantages over the independent lag model (it is easier and more intuitive to add additional explanatory variables to the joint model), differencing clearly has an important role to play in temporal analysis of EHR data. Incorporating rates of change must typically be done intentionally within any machine learning framework, including deep learning, either by pre-processing the features or by choosing model structures that learn temporal feature representations as linear combinations of neighboring sequential elements.

*Opportunities for revealing finer temporal structure in EHR data*
It is also important to note that lagged coefficients from these analyses contain information far more rich than the evaluated classifications (increasing, none, or decreasing physiologic responses). The trajectories of lagged coefficients (as seen in Figure 2, c.f. Figures 6-8 in [21], c.f. Figure 2 in [22]) can shed light on temporal dynamics and important time scales of the physiologic and/or health care process, rather than merely indicate the presence of an effect. We originally wanted to also evaluate these methods for their ability to detect finer temporal associations, but challenges remain for creating a reliable gold standard upon which to base validations of more complex insights, such as the rate or magnitude of a drug's physiologic effect (trustworthy quantitative information of this type does not exist for most cases). With sufficient validation, properly tuned lagged linear methods may eventually become useful for discovering novel associations in EHR data.

*Implications for comparing machine learning methods*
Most of the tested method combinations failed (AUROC=0.5), indicating that these choices are critically important. We observe that, for the same machine-learning algorithm, differences in preprocessing and experimental setup result in a range of AUROC from 0.5 to 0.8. Therefore, the choice of an overall algorithm (regression, support vector machines, neural networks, decision trees, etc.) is just one factor that could affect results, and researchers need to be mindful of this not only when performing experimental comparisons of algorithms, but also when presenting the results of these comparisons. While sophisticated machine-learning techniques aid learning of data representation, the structure for these models is still often selected based on certain hypotheses about how the data might be best represented. Our results suggest that data representations, either pre-processed or learned, should look like sequence time, and, most likely, contain information about the differences between successive measurements and normalize values across patients in the data set. Preprocessing conditions may have different effects on different methods, so a variety of these conditions ought to be rigorously tested, compared, and reported. The combination of pre-processing methodology and choice of gold standards could have large effects on machine learning evaluations, and it is likely that confidence intervals normally reported in machine learning studies fail to include the uncertainty related to these choices.

*Implications for reproducibility of observational studies*
Our evaluation pipeline is an important part of reproducible observational research, allowing researchers to quantify the impact of the various modeling choices made throughout the research process. Thorough comparisons across wider ranges of methods and source data are critical for advancing our ability to trust what we can learn from the EHR. The Observational Health Data Science and Informatics (OHDSI) consortium provides a common data model and a research community dedicated to such reproducible and generalizable advancements, and we aim to expand our pipeline into an OHDSI-compatible, open-source repository.

*How to choose the right method*
We have demonstrated the value of rigorous, systematic perturbations to chosen methods, and we encourage readers to perform similar evaluations in their own research contexts. However, we also hope that our results are somewhat generalizable to time-series analyses of medical data. We have found sequence time to provide a large, significant performance boon, and strongly recommend that researchers in similar domains consider re-indexing their time-series according to sequences. For lagged linear analysis, we recommend using either a simple independent lag model (without differencing) or a joint autoregressive model with differencing (recall that differencing corrupted the signals from the independent lag model). In general, we recommend performing differencing in accordance with results from statistical tests of unit root presence, like augmented Dickey-Fuller [34]. While we identified no statistical difference between the joint AR model with differencing and the independent lag model without differencing, qualitative assessment (e.g. see supplementary figures 1-7) suggests that the joint AR model provides finer resolution of temporal dynamics of physiologic process. In addition, even when the joint AR and independent lag models return the same drug-effect classifications, the joint AR model appears to be more robustly representative of the classification (e.g. supplementary figure 1, where it more clearly depicts that amphotericin B has no effect on total creatine kinase). These qualitative inspections cause us to favor the joint autoregressive model with differencing. Intra-patient normalization had no statistically significant effect in our cohort, but we recommend its continued usage, because a) it has been shown to improve performance in similar studies [22], and b) it did not create any disadvantage in our current study. We did not observe any useful effect from our experimental choice of windowing, and recommend readers select none or constant window functions as opposed our experimental choice. However, we encourage researchers to more thoroughly investigate appropriate windowing functions for EHR data, and insist that this be done in combination with other potential methodological choices, as there may be unexpected method-dependent dependencies. By studying the impact of methodological variations alone and in concert with each other, we can improve model performance and help make research results more generalizable and implementable for researchers.

## LIMITATIONS
This study was performed at a single academic medical center, and its findings may not generalize to different sources of medical record data. The gold standards are subject to existing, accessible knowledge. The selected method for classifying lagged coefficients was not studied rigorously, and may possess unforeseen biases.

## CONCLUSIONS
We used lagged linear methods to detect physiologic drug effects in EHR data. We used two clinical gold standards and a bootstrap methodology to evaluate the reliance of lagged methods on combinations of methodological perturbations. We observed important statistically significant improvements from particular combinations of temporal re-parameterization, time-series differencing, and regression model choice. We expect that these steps will play an important role in revealing fine temporal structure from EHR data, and we recognize the overarching importance of systematic comparison of machine learning methods under a broad range of pre-processing scenarios.

## ACKNOWLEDGMENTS
This work was supported by the National Library of Medicine grant number R01 LM006910.

**TABLES**

| Drug | Laboratory Measurement | Knowledge-base derived gold standard | Expert-curated gold standard |
|---|---|---|---|
| Allopurinol | Total Creatine Kinase | 1 | 0 |
| Allopurinol | Creatinine | 1 | 1 |
| Allopurinol | Potassium | 0 | 0 |
| Allopurinol | Hemoglobin | -1 | 0 |
| Amphotericin B | Total Creatine Kinase | 0 | 0 |
| Amphotericin B | Creatinine | 1 | 1 |
| Amphotericin B | Potassium | -1 | -1 |
| Amphotericin B | Hemoglobin | -1 | -1 |
| Furosemide | Total Creatine Kinase | 0 | 0 |
| Furosemide | Creatinine | 1 | 1 |
| Furosemide | Potassium | -1 | -1 |
| Furosemide | Hemoglobin | -1 | 1 |
| Ibuprofen | Total Creatine Kinase | 0 | 0 |
| Ibuprofen | Creatinine | 1 | 0 |
| Ibuprofen | Potassium | 1 | 0 |
| Ibuprofen | Hemoglobin | -1 | -1 |
| Simvastatin | Total Creatine Kinase | 1 | 1 |
| Simvastatin | Creatinine | 1 | 0 |
| Simvastatin | Potassium | 1 | 0 |
| Simvastatin | Hemoglobin | -1 | 0 |
| Spironolactone | Total Creatine Kinase | 0 | 0 |
| Spironolactone | Creatinine | 1 | 1 |
| Spironolactone | Potassium | 1 | 1 |
| Spironolactone | Hemoglobin | -1 | 1 |
| Warfarin | Total Creatine Kinase | 0 | 0 |
| Warfarin | Creatinine | 0 | 0 |
| Warfarin | Potassium | 0 | 0 |
| Warfarin | Hemoglobin | -1 | -1 |

Table 1. Clinical gold standards for expected drug effects.

# FIGURES

**Fig 1. Timeline Construction.** We performed a linear temporal interpolation in order to align sparse, asynchronous measurements and events. For every time point where there was a value (lab, drug concept, or context (i.e. inpatient admission)), the values of each other variable at that time point were interpolated as the clock-time weighted mean of the preceding and succeeding value of each respective variable.

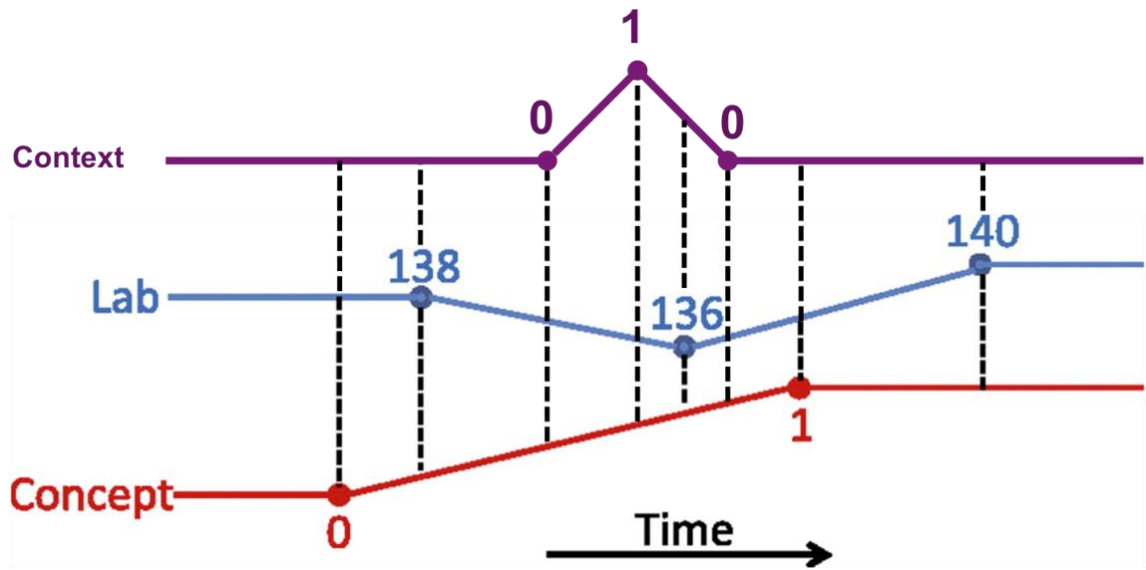

**Figure 2.** Experimental Design Overview

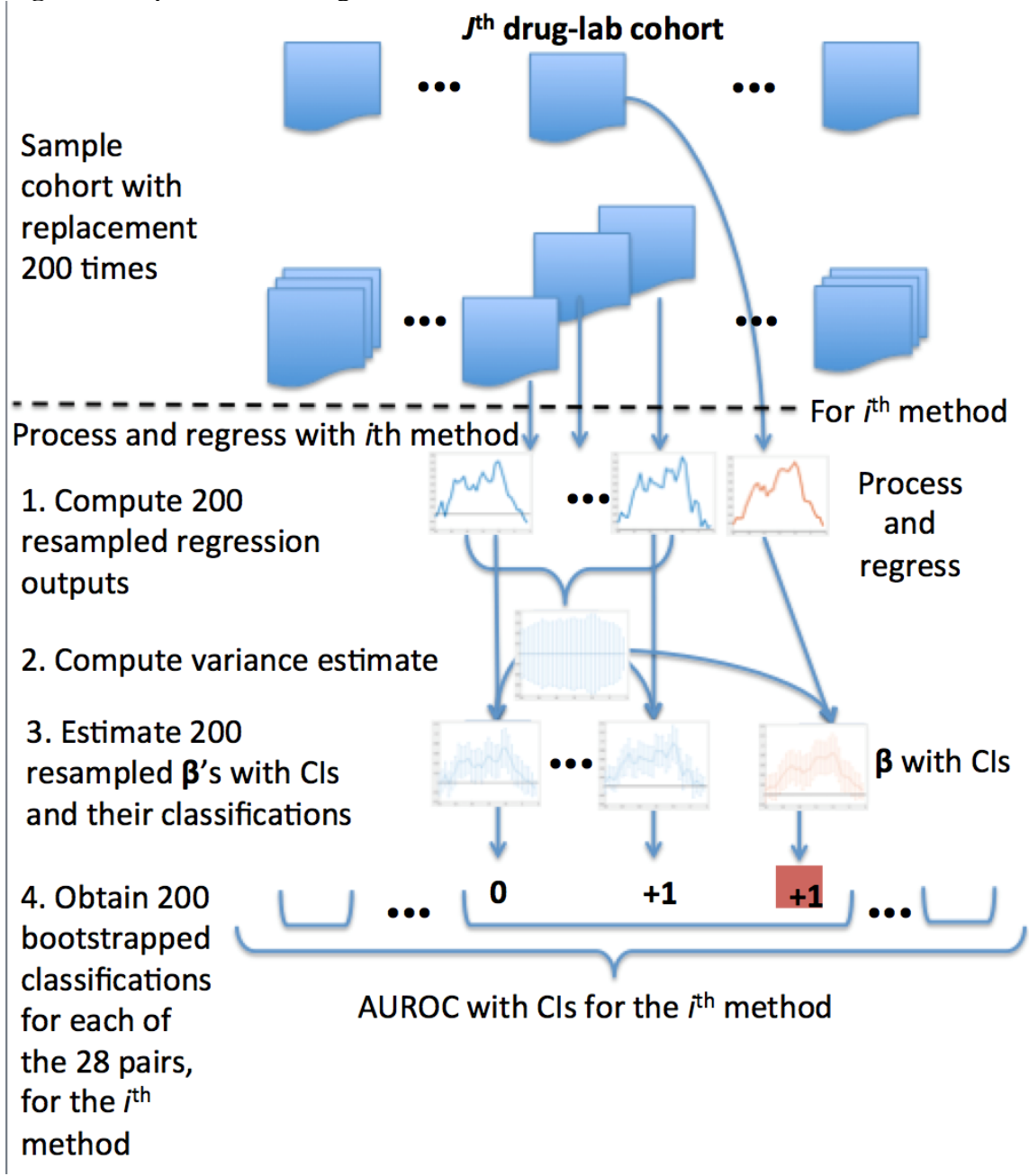

**Figure 3.** Signal quality is noticeably affected by combinations of methodological choices, especially temporal parameterizations, differencing, and lag model type; here we vary these 3 dimensions, and fix the remaining 3 using intra-patient normalization, no binning, and no additional context variables. Here, we expect Amphotericin B to increase Creatinine (hence, blue should be significantly above zero) and Amphotericin B to decrease Potassium (hence, red should be significantly below zero). The figures demonstrate that sequence-time is often a necessary, singular choice: figure 3a, which uses sequence time, produces the expected result, whereas figure 3b shows a non-significant noise pattern; the methods used in these figures differ only by their treatment of temporal parameterization. The figures also demonstrate that methods must be combined carefully—figure 3a combines differencing with the joint AR model, and produces expected patterns, whereas figure 3c uses an identical method, but omits differencing, and produces a non-significant noise pattern. However, pairing the independent lag model without differencing appears to reconstruct the signal, albeit with less significance than fig 3a.

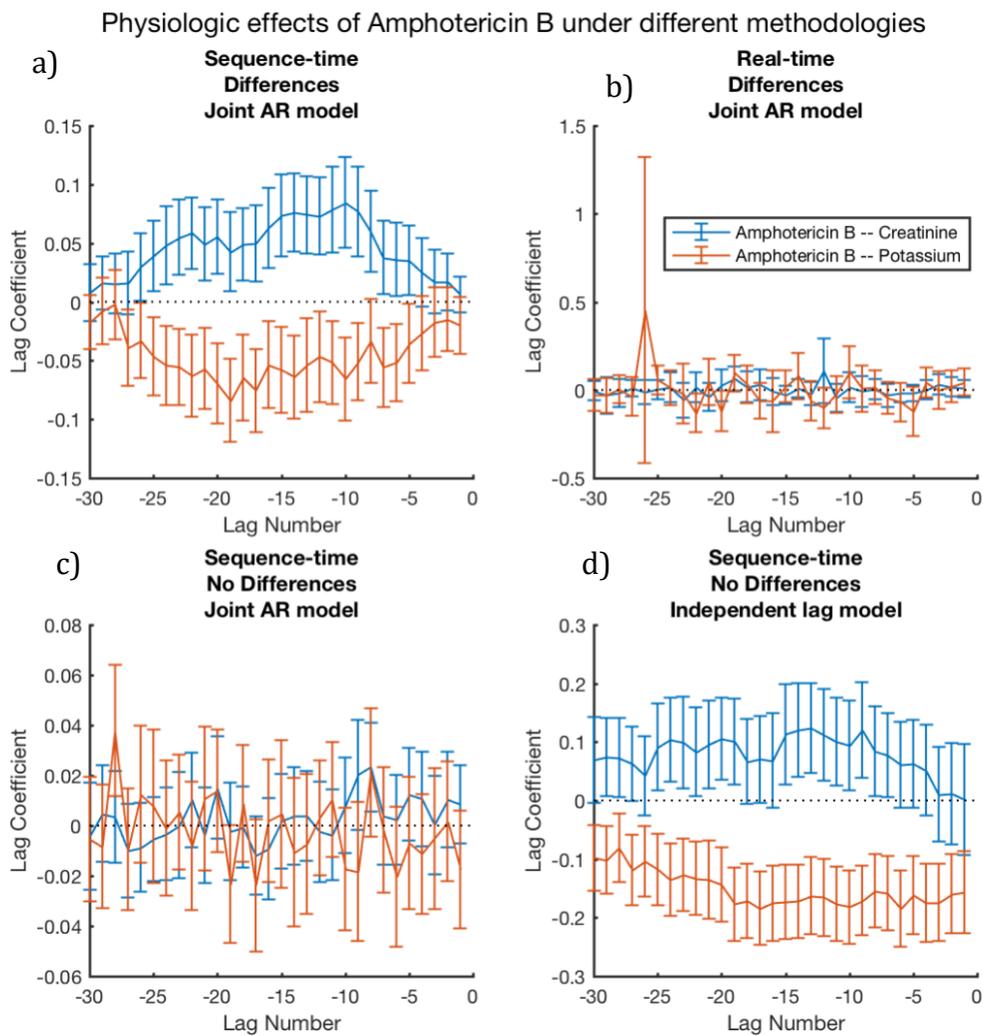

**Figure 4.**
This figure displays AUROC confidence intervals for each of $2^6$ methodological combinations. AUROCs are ordered from top-to-bottom in descending order of AUROC from the expert-curated gold standard. The heatmap on the left indicates the presence (tan) or absence (blue) of each of the 6 method variables for each plotted AUROC. For example, the top AUROC method (according to the clinically-curated gold standard) used sequence-time, no binning, intra-patient normalization, differencing, no additional context variable, and a joint AR model. Note that these results are enumerated explicitly in Supplementary Table 1.

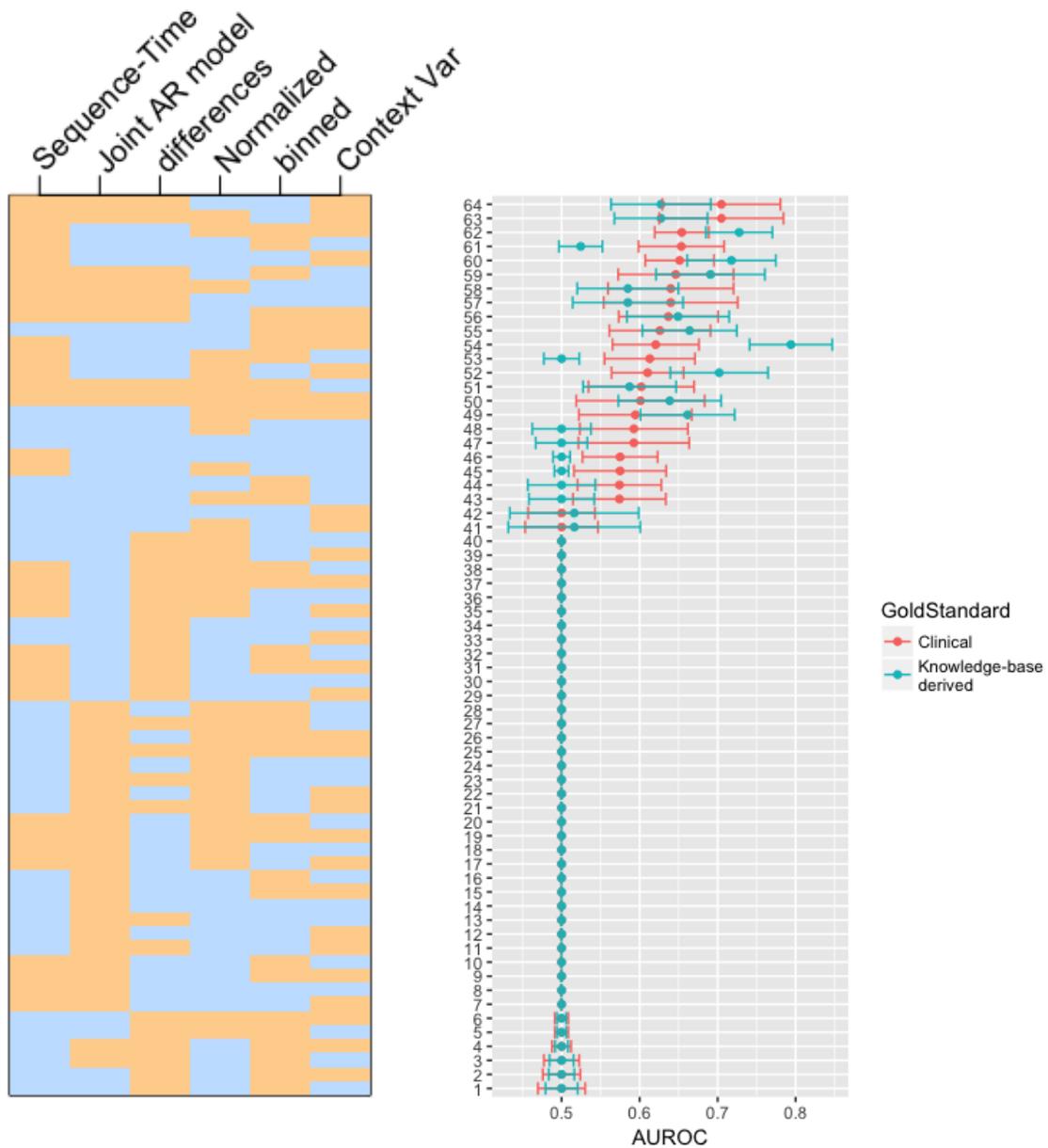

**Figure 5.**
Here we plot the correlation (Pearson correlation coefficient 0.759, p=3.7e-13) between AUROCs computed using clinically curated and knowledge-base derived gold standards. Error bars for each AUROC couple are 95% Confidence Intervals computed using a bootstrap resampling. We observe that the two gold standards, despite significant disagreements (Table 1), ultimately provide evaluations with reasonable similarity. This result instills a confidence in both gold standards that could not be achieved with a single gold standard.

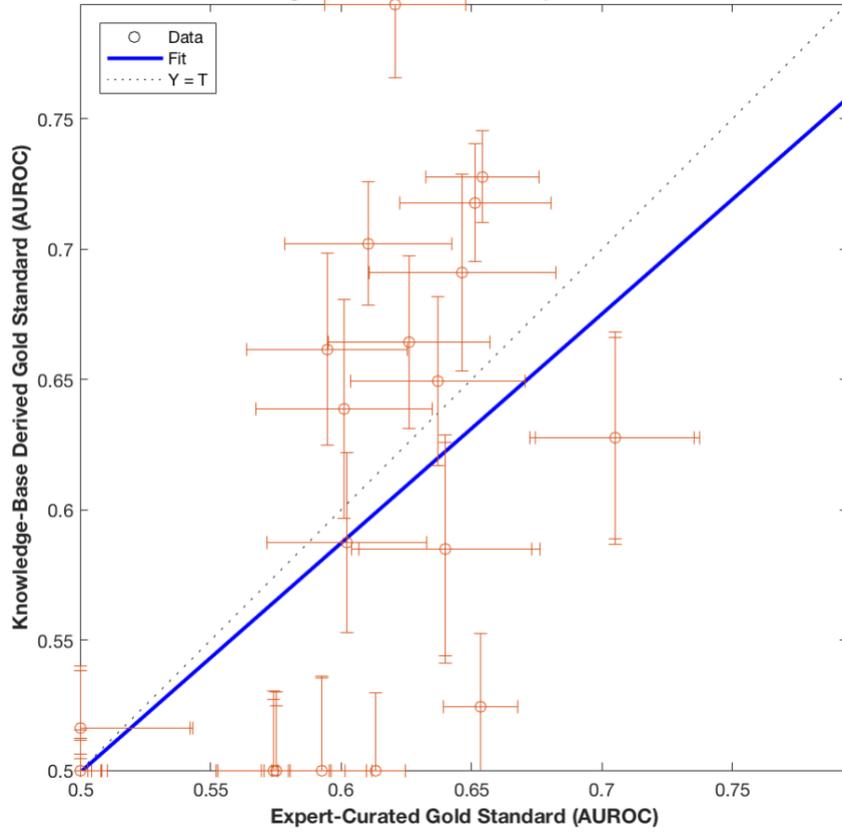

# SUPPLEMENTARY MATERIALS

## Supplementary Table 1

| Time | Binned | Normalized | Difference | Context variables | Estimation | Expert-curated Gold Standard AUROC | Expert-curated Gold Standard AUROC SD | Knowledge-base derived Gold Standard AUROC | Knowledge-base derived Gold Standard AUROC SD |
|---|---|---|---|---|---|---|---|---|---|
| Sequence | No | Yes | Yes | Yes | Joint AR | 0.705 | 0.04 | 0.627 | 0.03 |
| Sequence | No | Yes | Yes | No | Joint AR | 0.705 | 0.04 | 0.627 | 0.03 |
| Sequence | Yes | Yes | No | Yes | Independent | 0.654 | 0.02 | 0.728 | 0.02 |
| Sequence | Yes | No | No | No | Independent | 0.654 | 0.03 | 0.524 | 0.01 |
| Sequence | No | Yes | No | No | Independent | 0.651 | 0.02 | 0.718 | 0.03 |
| Sequence | Yes | No | Yes | No | Joint AR | 0.646 | 0.04 | 0.691 | 0.04 |
| Sequence | No | No | Yes | No | Joint AR | 0.640 | 0.04 | 0.585 | 0.04 |
| Sequence | No | No | Yes | Yes | Joint AR | 0.640 | 0.04 | 0.585 | 0.03 |
| Sequence | Yes | Yes | Yes | No | Joint AR | 0.637 | 0.03 | 0.649 | 0.03 |
| Real | Yes | Yes | No | No | Independent | 0.626 | 0.03 | 0.664 | 0.03 |
| Sequence | Yes | Yes | No | No | Independent | 0.621 | 0.03 | 0.794 | 0.03 |
| Sequence | Yes | No | No | Yes | Independent | 0.613 | 0.03 | 0.500 | 0.01 |
| Sequence | No | Yes | No | Yes | Independent | 0.610 | 0.02 | 0.702 | 0.03 |
| Sequence | Yes | No | Yes | Yes | Joint AR | 0.602 | 0.03 | 0.587 | 0.03 |
| Sequence | Yes | Yes | Yes | Yes | Joint AR | 0.601 | 0.04 | 0.639 | 0.03 |
| Real | Yes | Yes | No | Yes | Independent | 0.595 | 0.04 | 0.662 | 0.03 |
| Real | No | No | No | No | Independent | 0.593 | 0.04 | 0.500 | 0.02 |
| Real | No | No | No | Yes | Independent | 0.593 | 0.04 | 0.500 | 0.02 |
| Sequence | No | No | No | No | Independent | 0.575 | 0.02 | 0.500 | 0.01 |
| Sequence | No | No | No | Yes | Independent | 0.575 | 0.03 | 0.500 | 0.00 |
| Real | Yes | No | No | No | Independent | 0.574 | 0.03 | 0.500 | 0.02 |
| Real | Yes | No | No | Yes | Independent | 0.574 | 0.03 | 0.500 | 0.02 |
| Real | No | Yes | No | No | Independent | 0.500 | 0.02 | 0.516 | 0.04 |
| Real | No | Yes | No | Yes | Independent | 0.500 | 0.02 | 0.516 | 0.04 |
| Sequence | No | Yes | No | No | Joint AR | 0.500 | 0.00 | 0.500 | 0.00 |
| Sequence | No | No | No | No | Joint AR | 0.500 | 0.00 | 0.500 | 0.00 |
| Sequence | Yes | Yes | No | No | Joint AR | 0.500 | 0.00 | 0.500 | 0.00 |
| Sequence | Yes | No | No | No | Joint AR | 0.500 | 0.00 | 0.500 | 0.00 |
| Real | No | Yes | Yes | No | Joint AR | 0.500 | 0.00 | 0.500 | 0.00 |
| Real | No | Yes | No | No | Joint AR | 0.500 | 0.00 | 0.500 | 0.00 |
| Real | No | No | Yes | No | Joint AR | 0.500 | 0.00 | 0.500 | 0.00 |
| Real | No | No | No | No | Joint AR | 0.500 | 0.00 | 0.500 | 0.00 |
| Real | Yes | Yes | Yes | No | Joint AR | 0.500 | 0.01 | 0.500 | 0.00 |
| Real | Yes | Yes | No | No | Joint AR | 0.500 | 0.00 | 0.500 | 0.00 |
| Real | Yes | No | Yes | No | Joint AR | 0.500 | 0.01 | 0.500 | 0.01 |
| Real | Yes | No | No | No | Joint AR | 0.500 | 0.00 | 0.500 | 0.00 |
| Sequence | No | Yes | No | Yes | Joint AR | 0.500 | 0.00 | 0.500 | 0.00 |
| Sequence | No | No | No | Yes | Joint AR | 0.500 | 0.00 | 0.500 | 0.00 |
| Sequence | Yes | Yes | No | Yes | Joint AR | 0.500 | 0.00 | 0.500 | 0.00 |
| Sequence | Yes | No | No | Yes | Joint AR | 0.500 | 0.00 | 0.500 | 0.00 |
| Real | No | Yes | Yes | Yes | Joint AR | 0.500 | 0.00 | 0.500 | 0.00 |
| Real | No | Yes | No | Yes | Joint AR | 0.500 | 0.00 | 0.500 | 0.00 |
| Real | No | No | Yes | Yes | Joint AR | 0.500 | 0.00 | 0.500 | 0.00 |
| Real | No | No | No | Yes | Joint AR | 0.500 | 0.00 | 0.500 | 0.00 |
| Real | Yes | Yes | Yes | Yes | Joint AR | 0.500 | 0.00 | 0.500 | 0.00 |
| Real | Yes | Yes | No | Yes | Joint AR | 0.500 | 0.00 | 0.500 | 0.00 |
| Real | Yes | No | Yes | Yes | Joint AR | 0.500 | 0.00 | 0.500 | 0.00 |
| Real | Yes | No | No | Yes | Joint AR | 0.500 | 0.00 | 0.500 | 0.00 |
| Sequence | No | Yes | Yes | No | Independent | 0.500 | 0.00 | 0.500 | 0.00 |
| Sequence | No | No | Yes | No | Independent | 0.500 | 0.00 | 0.500 | 0.00 |
| Sequence | Yes | Yes | Yes | No | Independent | 0.500 | 0.00 | 0.500 | 0.00 |
| Sequence | Yes | No | Yes | No | Independent | 0.500 | 0.00 | 0.500 | 0.00 |
| Real | No | Yes | Yes | No | Independent | 0.500 | 0.00 | 0.500 | 0.00 |
| Real | No | No | Yes | No | Independent | 0.500 | 0.00 | 0.500 | 0.00 |
| Real | Yes | Yes | Yes | No | Independent | 0.500 | 0.01 | 0.500 | 0.01 |
| Real | Yes | No | Yes | No | Independent | 0.500 | 0.02 | 0.500 | 0.01 |
| Sequence | No | Yes | Yes | Yes | Independent | 0.500 | 0.00 | 0.500 | 0.00 |
| Sequence | No | No | Yes | Yes | Independent | 0.500 | 0.00 | 0.500 | 0.00 |
| Sequence | Yes | Yes | Yes | Yes | Independent | 0.500 | 0.00 | 0.500 | 0.00 |
| Sequence | Yes | No | Yes | Yes | Independent | 0.500 | 0.00 | 0.500 | 0.00 |
| Real | No | Yes | Yes | Yes | Independent | 0.500 | 0.00 | 0.500 | 0.00 |
| Real | No | No | Yes | Yes | Independent | 0.500 | 0.00 | 0.500 | 0.00 |
| Real | Yes | Yes | Yes | Yes | Independent | 0.500 | 0.00 | 0.500 | 0.00 |
| Real | Yes | No | Yes | Yes | Independent | 0.500 | 0.00 | 0.500 | 0.00 |

Supplementary Table 1 lists the evaluation metrics (AUC with standard deviation with respect to each gold standard) for each of the 64 methodological combinations.

**Supplementary Table 2**

| Allopurinol |
|---|
| CPMC DRUG: ALLOPURINOL 100 MG TAB |
| CPMC DRUG: ALLOPURINOL 300 MG TAB |
| CPMC DRUG: UD ALLOPURINOL 100 MG TAB |
| CPMC DRUG: UD ALLOPURINOL 300 MG TAB |
| ALLOPURINOL |
| ALLOPURINOL PREPARATIONS |
| ALLOPURINOL 100 MG TABLET |
| ALLOPURINOL 300 MG TABLET |
| ALLOPURINOL TABLETS |
| CERNER DRUG: ALLOPURINOL TAB 300 MG |
| CERNER DRUG: ALLOPURINOL SUSP 5 MG/ML |
| CERNER DRUG: ALLOPURINOL TAB 100 MG |
| CERNER DRUG: ALLOPURINOL INJ 500 MG |
| CERNER DRUG: ALLOPURINOL PO SUSP 20 MG/ML EXT |
| ALLOPURINOL SUSP 5 MG/ML |
| ALLOPURINOL INJ 500 MG |
| ALLOPURINOL 20 MG/ML PO SUSPENSION |
| CERNER DRUG: ALLOPURINOL IV SYR 6 MG/ML D5W |
| CERNER DRUG: ALLOPURINOL IV SYR 6 MG/ML NS |
|  |
| **Amphotericin B** |
| CPMC DRUG: FUNGIZONE 50 MG VIAL |
| AMPHOTERICIN B |
| AMPHOTERICIN B PREPARATIONS |
| CPMC DRUG: AMPHOTERICIN B 50MG FOR AEROSO |
| CPMC DRUG: AMPHOTERICIN B (FUNGIZONE) |
| CPMC DRUG: AMPHOTERICIN 3% CREAM 20 GM |
| CPMC DRUG: .AMPHOTERICIN B (FUNGIZONE) |
| CPMC DRUG: AMPHOTERICIN B LIPID 5MG/ML IN |
| AMPHOTERICIN B 3 % |
| AMPHOTERICIN B 0 IDA VIAL |
| AMPHOTERICIN B 5 MG/ML |
| CPMC DRUG: FUNGIZONE OR SUS 100MG/ML 24ML |
| CPMC DRUG: .AMPHOTERICIN B LIPID 5MG/ML |
| CERNER DRUG: AMPHOTERICIN B LIPOSOME IVPB *R* |
| CERNER DRUG: AMPHOTERICIN B SUSP 100 MG/ML 24 ML |
| CERNER DRUG: AMPHOTERICIN B LIPID 5 MG/ML INJ *NF* |
| CERNER DRUG: AMPHOTERICIN B LIPID 2 MG/ML INJ SYR *NF* |
| CERNER DRUG: AMPHOTERICIN B INJ 50 MG |
| CERNER DRUG: AMPHOTERICIN B APPROVAL |
| CERNER DRUG: AMPHOTERICIN B - 0.1 MG/ML INJ SYR *R* |
| CERNER DRUG: AMPHOTERICIN B EXT OPHT 5 MCG/0.1 ML INJ 0.3 ML |
| CERNER DRUG: AMPHOTERICIN B LIPID APPROVAL |
| CERNER DRUG: AMPHOTERICIN B LIPOSOME APPROVAL |
| CERNER DRUG: AMPHOTERICIN B LIPOSOME *APPROVAL* |
| CERNER DRUG: AMPHOTERICIN B LIPOSOME APPROVAL (2) |
| AMPHOTERICIN B SUSP 100 MG/ML |
| AMPHOTERICIN B 1 MG/ML |
| AMPHOTERICIN B 2 MG/ML |
| AMPHOTERICIN B 0.1 MG/ML |
| AMPHOTERICIN B 0.25 MG/ML |
| AMPHOTERICIN B LIPID PREPARATIONS |
| AMPHOTERICIN B LIPOSOME PREPARATIONS |
| CERNER DRUG: AMPHOTERICIN B LIPID NEBULIZER 5 MG/ML *NF* |
| CERNER DRUG: AMPHO B LIPOSOME IV SY 2 MG/ML D5W |
| CERNER DRUG: AMPHOTERICIN B SOLN 0.1 MG/ML 10ML |
| CERNER DRUG: AMPHOTERICIN B - 0.5 MG/ML INJ SYR *R* |
| CERNER DRUG: AMPHO B DEOXY OPH DROP 1.5MG/ML 10M |
| CERNER DRUG: AMPHO B DEOXY 0.1 MG/ML CNS *R* |
| CERNER DRUG: AMPHOTERICIN B INTRATHECAL APPROVAL |
| CERNER DRUG: AMPHOTERICIN B LIPOSOMAL 50 MG INJ *R* |
| CERNER DRUG: AMPHOTERICIN B LIPOSOMAL NEBULIZER 25 MG/6 ML SOLN |
| |
| **Furosemide** |
| FUROSEMIDE PREPARATIONS |
| FUROSEMIDE |
| CPMC DRUG: FUROSEMIDE 20 MG TAB |
| CPMC DRUG: FUROSEMIDE 40 MG TAB |
| CPMC DRUG: LASIX 10 MG/ML 10 ML AMP |
| CPMC DRUG: LASIX 10 MG/ML 2 ML AMP |
| CPMC DRUG: LASIX SOLN 10 MG/ML 60 ML |
| CPMC DRUG: UD FUROSEMIDE 20 MG TAB |
| CPMC DRUG: UD FUROSEMIDE 40 MG TAB |
| CPMC DRUG: LASIX 10 MG/ML 2 ML INJ |
| CPMC DRUG: FUROSEMIDE 10 MG/ML SOL |
| CPMC DRUG: FUROSEMIDE 10 MG/ML SOL 60 ML |
| CPMC DRUG: FUROSEMIDE 10 MG/ML 2 ML INJ |
| CPMC DRUG: FUROSEMIDE 100 MG/10 ML INJ |
| CPMC DRUG: .LASIX 10 MG/ML 2 ML INJ |
| CPMC DRUG: .FUROSEMIDE 20 MG TAB |
| OPERATING ROOM MEDICATION: FUROSEMIDE |
| FUROSEMIDE 80MG TABLET |
| FUROSEMIDE 10 MG/ML |
| FUROSEMIDE 20 MG TABLET |
| FUROSEMIDE 40 MG TABLET |
| FUROSEMIDE TABLETS |
| CPMC DRUG: UD LASIX 40 MG TAB |
| CERNER DRUG: FUROSEMIDE TAB 40 MG |
| CERNER DRUG: FUROSEMIDE SOLN 10 MG/ML 120 ML B |
| CERNER DRUG: FUROSEMIDE ELIX 40 MG/5 ML DU |
| CERNER DRUG: FUROSEMIDE INJ 10 MG/ML 4 ML |
| CERNER DRUG: FUROSEMIDE SOLN 10 MG/ML 60 ML B |
| CERNER DRUG: FUROSEMIDE NEB *IND* 2 ML |
| CERNER DRUG: FUROSEMIDE TAB 20 MG |
| CERNER DRUG: FUROSEMIDE INJ 10 MG/ML 2 ML |
| CERNER DRUG: FUROSEMIDE IVPB COMPOUND |
| CERNER DRUG: FUROSEMIDE IV SY 5 MG/ML NS |
| CERNER DRUG: FUROSEMIDE IV SY 10 MG/ML |
| CERNER DRUG: FUROSEMIDE INJ 10 MG/ML 10 ML |
| CERNER DRUG: FUROSEMIDE INFUSION 10 MG/ML |
| CERNER DRUG: FUROSEMIDE TAB 80 MG |
| CERNER DRUG: FUROSEMIDE PO SOLN 10 MG/ML EXT |
| FUROSEMIDE ELIX 8 MG/ML |
| FUROSEMIDE IV 5 MG/ML |
| FUROSEMIDE 80 MG TABLET |
| CERNER DRUG: FUROSEMIDE IV SY 1 MG/ML NS |
| CERNER DRUG: FUROSEMIDE 1 MG/ML SYRINGE |
| CERNER DRUG: FUROSEMIDE 5 MG/ML SYRINGE |
| CERNER DRUG: FUROSEMIDE 0.5 MG/ML SYRINGE |
| CERNER DRUG: FUROSEMIDE 10 MG/ML SYRINGE |
| CERNER DRUG: FUROSEMIDE 10 MG/ML INJ SYR |
| CERNER DRUG: FUROSEMIDE 100 MG/100 ML NS |
| CERNER DRUG: FUROSEMIDE 40 MG/4 ML SOLN UD |
| |
| **Simvastatin** |
| SIMVASTATIN PREPARATIONS |
| CPMC DRUG: SIMVASTATIN W/LACTOSE 10MG TAB |
| CPMC DRUG: UD SIMVASTATIN 5 MG TAB |
| CPMC DRUG: UD SIMVASTATIN 10 MG TAB |
| CPMC DRUG: SIMVASTATIN 20 MG TAB |
| CPMC DRUG: SIMVASTATIN 40 MG TAB |
| SIMVASTATIN TABLETS |
| SIMVASTATIN 10 MG TABLET |
| SIMVASTATIN 20 MG TABLET |
| SIMVASTATIN 40 MG TABLET |
| SIMVASTATIN 5 MG TABLET |
| SIMVASTATIN |
| CERNER DRUG: SIMVASTATIN TAB 20 MG |
| CERNER DRUG: SIMVASTATIN TAB 40 MG |
| CERNER DRUG: SIMVASTATIN TAB 10 MG |
| CERNER DRUG: SIMVASTATIN TAB 5 MG |
| SIMVASTATIN 80MG TABLET |
| CERNER DRUG: SIMVASTATIN TAB 80 MG |
| |
| **Spironolactone** |
| CPMC DRUG: ALDACTONE 25 MG TAB |
| CPMC DRUG: UD ALDACTONE 25 MG TAB |
| SPIRONOLACTONE |
| SPIRONOLACTONE PREPARATIONS |
| CPMC DRUG: SPIRONOLACTONE 25 MG TAB |
| CPMC DRUG: UD SPIRONOLACTONE 25 MG TAB |
| CPMC DRUG: SPIRONOLACTONE SUS 5 MG/ML |
| SPIRONOLACTONE 5 MG/ML |
| SPIRONOLACTONE 25 MG TABLET |
| SPIRONOLACTONE TABLETS |
| SPIRONOLACTONE 100 MG TABLET |
| SPIRONOLACTONE 50 MG TABLET |
| CERNER DRUG: SPIRONOLACTONE PO SUSP 5 MG/ML EXT |
| CERNER DRUG: SPIRONOLACTONE-HCTZ *NF* 25 MG |
| CERNER DRUG: SPIRONOLACTONE TAB 25 MG |
| CERNER DRUG: SPIRONOLACTONE TAB 100 MG *NF* |
| |
| **Ibuprofen** |
| CPMC DRUG: MOTRIN 400 MG TAB |
| CPMC DRUG: MOTRIN 600 MG TAB |
| CPMC DRUG: UD MOTRIN 400 MG TAB |
| CPMC DRUG: UD MOTRIN 600 MG TAB |
| IBUPROFEN |
| IBUPROFEN PREPARATIONS |
| CPMC DRUG: PEDIAPROFEN 100 MG/5 ML 120 ML |
| CPMC DRUG: UD IBUPROFEN 600 MG TAB |
| CPMC DRUG: UD IBUPROFEN 400 MG TAB |
| IBUPROFEN 100 MG/ML |
| IBUPROFEN 400 MG TABLET |
| IBUPROFEN 600 MG TABLET |
| IBUPROFEN 800 MG TABLET |
| IBUPROFEN TABLETS |
| IBUPROFEN 100 MG TABLET |
| IBUPROFEN 200 MG TABLET |
| IBUPROFEN 300 MG TABLET |
| IBUPROFEN CHEWABLE TABLETS |
| IBUPROFEN 100 MG CHEWABLE TABLET |
| IBUPROFEN 50 MG CHEWABLE TABLET |
| IBUPROFEN CAPSULES |
| IBUPROFEN 200 MG CAPSULE |
| CPMC DRUG: ADVIL CAPLETS |
| CERNER DRUG: IBUPROFEN TAB 400 MG |
| CERNER DRUG: IBUPROFEN TAB 200 MG |
| CERNER DRUG: IBUPROFEN TAB 600 MG |
| CERNER DRUG: IBUPROFEN SUSP 100 MG/5 ML 120ML B |
| CERNER DRUG: IBUPROFEN PO SUSP 20 MG/ML EXTEMP |
| IBUPROFEN SUSP 20 MG/ML |
| CERNER DRUG: IBUPROFEN SUSP 100 MG/5 ML UD |
| CERNER DRUG: IBUPROFEN LYSINE INJ 20 MG/2 ML *R* |
| IBUPROFEN, LYSINE SALT PREPARATIONS |
| LYSINE SALT OF IBUPROFEN |
| CERNER DRUG: IBUPROFEN LYSINE IV SY 5 MG/ML D5W |
| CERNER ME ORDER: ZZIBUPROFEN (ARUP) |
| CERNER ME DTA: IBUPROFEN |
| CERNER DRUG: IBUPROFEN INJ 100 MG/ML 8 ML *R* |
| CERNER DRUG: IBUPROFEN INJ 100 MG/ML 4 ML *R* |
| CERNER DRUG: IBUPROFEN IVPB *R* |
| |
| **Warfarin** |
| CPMC DRUG: COUMADIN 10 MG TAB |
| CPMC DRUG: COUMADIN 2 MG TAB |
| CPMC DRUG: COUMADIN 2.5 MG TAB |
| CPMC DRUG: COUMADIN 5 MG TAB |
| CPMC DRUG: UD COUMADIN 10 MG TAB |
| CPMC DRUG: UD COUMADIN 2 MG TAB |
| CPMC DRUG: UD COUMADIN 2.5 MG TAB |
| CPMC DRUG: UD COUMADIN 5 MG TAB |
| CPMC DRUG: COUMADIN 7.5 MG TAB |
| CPMC DRUG: UD WARFARIN SOD 1 MG TAB |
| CPMC DRUG: UD WARFARIN 2 MG TAB |
| WARFARIN 7.5 MG TABLET |
| WARFARIN 1 MG TABLET |
| WARFARIN 2 MG TABLET |
| WARFARIN 2.5 MG TABLET |
| WARFARIN 5 MG TABLET |
| WARFARIN 10 MG TABLET |
| WARFARIN TABLETS |
| WARFARIN SODIUM 4 MG TABLET |
| CERNER DRUG: WARFARIN TAB 1 MG |
| CERNER DRUG: WARFARIN TAB 10 MG *DNO* |
| CERNER DRUG: WARFARIN TAB 7.5 MG |
| CERNER DRUG: WARFARIN TAB 2 MG |
| CERNER DRUG: WARFARIN TAB 2.5 MG |
| CERNER DRUG: WARFARIN TAB 5 MG |
| CERNER DRUG: WARFARIN TAB 0.5 MG (PRODUCTION) |
| WARFARIN 0.5 MG TABLET |
| CERNER DRUG: WARFARIN 3 MG TAB |
| WARFARIN PREPARATIONS |
| |
| **Total Creatine Kinase** |
| SERUM CREATINE KINASE TEST |
| SERUM CREATINE PHOSPHOKINASE MEASUREMENT |
| SERUM TOTAL CREATINE KINASE TEST |
| SERUM CREATINE KINASE MEASUREMENT 2 |
| SERUM CREATINE KINASE TEST 2 |
| NEW CHEM20 PLASMA CREATINE KINASE TEST |
| CPMC LABORATORY TEST: OLD PLASMA CREATINE PHOSPHOKINASE |
| CPMC LABORATORY TEST: CK, TOTAL |
| PLASMA TOTAL CREATINE KINASE TEST |
| NYH LAB PROCEDURE: CREATINE KINASE |
| NYH LAB PROCEDURE: TOTAL CK (CK ISOENEZYME) |
| CPMC LABORATORY TEST: SERUM CREATINE KINASE |
| CPMC LABORATORY TEST: CREATINE KINASE(ALLEN) |
| CPMC LABORATORY TEST: CK,TOTAL |
| CPMC LABORATORY TEST: CK TOTAL |
| CERNER ME DTA: CREATINE KINASE |
| CERNER ME DTA: CK, TOTAL (QUEST 88001232) |
| |
| **Creatinine** |
| SERUM CREATININE MEASUREMENT |
| PRESBYTERIAN PLASMA CREATININE TEST |
| ALLEN PLASMA CREATININE MEASUREMENT |
| CHEM-7 CREATININE MEASUREMENT |
| SERUM CREATININE TESTS |
| SERUM CREATININE MEASUREMENT 2 |
| NEW CHEM-7 PLASMA CREATININE MEASUREMENT |
| CPMC LABORATORY TEST: OLD PLASMA CREATININE MEASUREMENT |
| CPMC LABORATORY TEST: CREATININE |
| NYH LAB PROCEDURE: CREATININE |
| NYH LAB PROCEDURE: PRE CREATININE |
| NYH LAB PROCEDURE: POST CREATININE |
| CPMC LABORATORY TEST: SERUM CREATININE MEASUREMENT |
| WHOLE BLOOD CREATININE TESTS |
| CPMC LABORATORY TEST: CREATININE, WHOLE BLOOD |
| CREATININE MANUALLY ENTERED BY HEALTH PROFESSIONAL |
| CPMC LABORATORY TEST: CREATININE WHOLE BLOOD |
| CPMC LABORATORY TEST: CREATININE WHOLE BLOOD (ALLEN) |
| CERNER ME DTA: CREATININE |
| NYH LAB PROCEDURE: CREATININE, W/B |
| CPMC LABORATORY TEST: CREATININE  (ISTAT) |
| CERNER ME DTA: CREATININE WHOLE BLOOD POC |
| CERNER ME DTA: CREATININE WB |
| CERNER ME DTA: CREATININE (QUEST) |
| CERNER ME DTA: CREATININE BGV |
| CERNER ME DTA: CREATININE BGA |
| CERNER ME DTA: CRE WB - EPOC |
| CERNER ME DTA: CREATININE (PLASMA) |
| INTRAVASCULAR CREATININE TEST |
| |
| **Potassium** |
| STAT WHOLE BLOOD POTASSIUM ION MEASUREMENT |
| PRESBYTERIAN WHOLE BLOOD POTASSIUM ION MEASUREMENT |
| SERUM POTASSIUM ION MEASUREMENT |
| PRESBYTERIAN PLASMA POTASSIUM ION TEST |
| ALLEN PLASMA POTASSIUM ION MEASUREMENT |
| ALLEN WHOLE BLOOD POTASSIUM ION MEASUREMENT |
| CHEM-7 POTASSIUM ION MEASUREMENT |
| WHOLE BLOOD POTASSIUM TESTS |
| SERUM POTASSIUM ION TESTS |
| SERUM POTASSIUM ION MEASUREMENT 2 |
| NEW CHEM-7 PLASMA POTASSIUM ION MEASUREMENT |
| CPMC LABORATORY TEST: OLD PLASMA POTASSIUM MEASUREMENT |
| CPMC LABORATORY TEST: POTASSIUM, WHOLE BLOOD |
| CPMC LABORATORY TEST: K WHOLE BLOOD |
| CPMC LABORATORY TEST: POTASSIUM(ALLEN) |
| OPERATING ROOM MISC LABS: K |
| NYH LAB PROCEDURE: POTASSIUM |
| NYH LAB PROCEDURE: POTASSIUM, PLASMA |
| NYH LAB PROCEDURE: POTASSIUM , W/B |
| CPMC LABORATORY TEST: SERUM POTASSIUM MEASUREMENT |
| PLASMA POTASSIUM TESTS |
| CPMC LABORATORY TEST: POTASSIUM ISTAT |
| POTASSIUM MANUALLY ENTERED BY HEALTH PROFESSIONAL |
| CPMC LABORATORY TEST: POTASSIUM WHOLE BLOOD |
| CPMC LABORATORY TEST: POTASSIUM WHOLE BLOOD (ALLEN) |
| CERNER ME DTA: POTASSIUM LEVEL |
| CERNER ME DTA: POTASSIUM WHOLE BLOOD POC |
| CERNER ME DTA: POTASSIUM WB |
| CERNER ME DTA: POTASSIUM PLASMA |
| CERNER ME DTA: POTASSIUM W/B - EPOC |
| CPMC LABORATORY TEST: POTASSIUM WHOLE BLOOD POC |
| CERNER ME DTA: POTASSIUM POC IL |
| CERNER ME DTA: POTASSIUM-TOTAL RBC (QUEST) |
| CERNER ME DTA: K POST |
| INTRAVASCULAR POTASSIUM TEST |
| |

| Hemoglobin |
|---|
| STAT WHOLE BLOOD HEMOGLOBIN MEASUREMENT |
| PRESBYTERIAN PATHOLOGY WHOLE BLOOD HEMOGLOBIN MEASUREMENT |
| PRESBYTERIAN CHEMISTRY WHOLE BLOOD HEMOGLOBIN MEASUREMENT |
| ALLEN WHOLE BLOOD HEMOGLOBIN TEST |
| ALLEN WHOLE BLOOD HEMOGLOBIN MEASUREMENT |
| RESPIRATORY BLOOD HEMOGLOBIN MEASUREMENT |
| STAT LABORATORY HEMOGLOBIN MEASUREMENT |
| CPMC LABORATORY TEST: HGB(C) |
| CPMC LABORATORY TEST: HEMOGLOBIN |
| CPMC LABORATORY TEST: HGB |
| CPMC LABORATORY TEST: MEASURED HEMOGLOBIN |
| CPMC LABORATORY TEST: TOTAL HEMOGLOBIN |
| OPERATING ROOM MISC LABS: HGB |
| NYH LAB PROCEDURE: TOTAL HEMOGLOBIN |
| NYH LAB PROCEDURE: HEMOGLOBIN |
| CPMC LABORATORY TEST: HEMOGLOBIN 59947 |
| NYH LAB PROCEDURE: HEMOGLOBIN W/B |
| CPMC LABORATORY TEST: HGB (M) |
| CPMC LABORATORY TEST: HGB (M) (ALLEN) |
| CPMC LABORATORY TEST: TOTAL HEMOGLOBIN (C) |
| CPMC LABORATORY TEST: HEMOGLOBIN, POC |
| CERNER ME DTA: HEMOGLOBIN |
| CERNER ME DTA: HEMOGLOBIN POC |
| CERNER ME DTA: HEMOGLOBIN TOTAL POC |
| CERNER ME DTA: HEMOGLOBIN WB |
| CERNER ME DTA: TOTAL HGB |
| CERNER ME DTA: HEMOGLOBIN W/B - EPOC |
| CPMC LABORATORY TEST: HEMOGLOBIN POC |
| CERNER ME DTA: TOTAL HGB POC IL (FOR CO-OX) |
| CERNER ME DTA: HGB-UNV |
| CERNER ME DTA: TOTAL HEMOGLOBIN POC IL |
| CERNER ME DTA: HEMOGLOBIN WHOLE BLOOD POC |
| CERNER ME DTA: HEMOGLOBIN (ARUP 2011401) |
| WHOLE BLOOD HEMOGLOBIN CONCENTRATION TESTS |

Supplementary Table 2 lists the names of all laboratory tests and drug orders extracted for the analysis.

**Supplementary Figure 1**

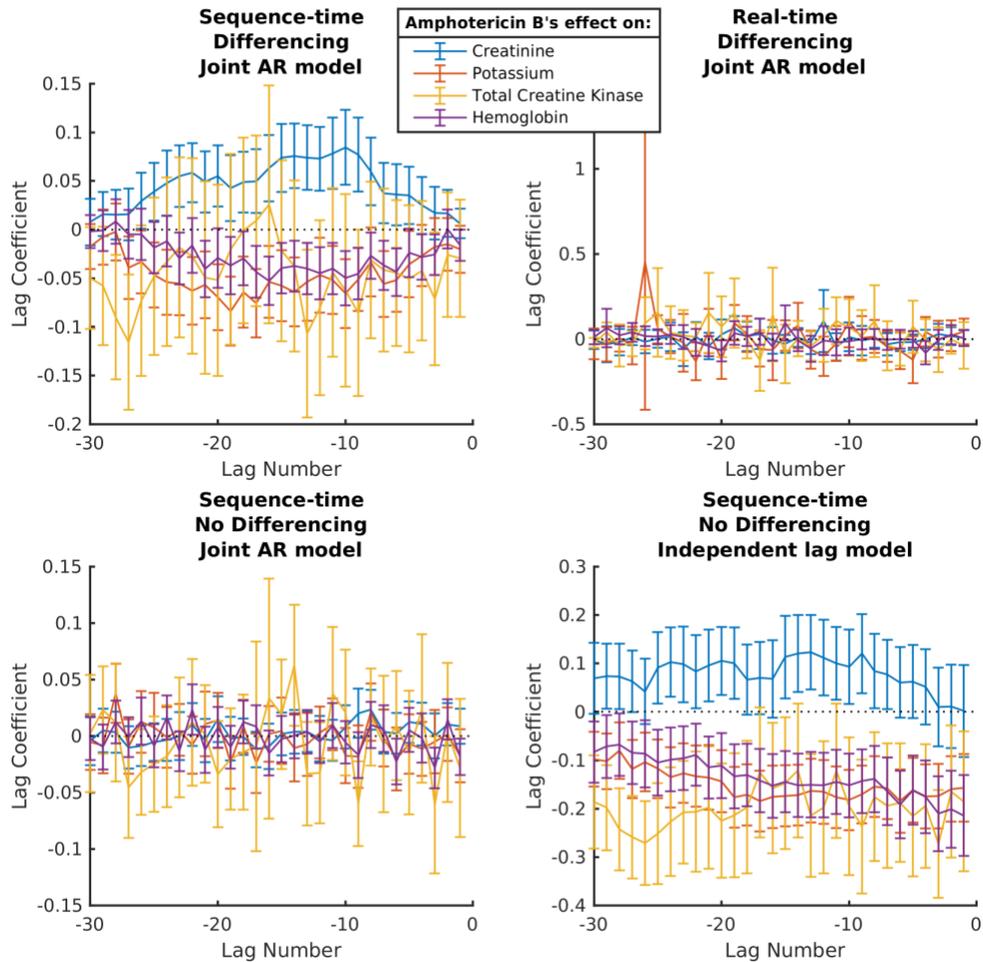

Lagged regression results from four different methodological variations are shown here for amphotericin B's effect on creatinine, potassium, total creatine kinase, and hemoglobin. Specifically, we show 4 illustrative combinations of temporal parameterization, differencing, and model choices; in all figures, intra-patient normalization was used, no windowing or binning was performed, and no additional context variable was used.

**Supplementary Figure 2**

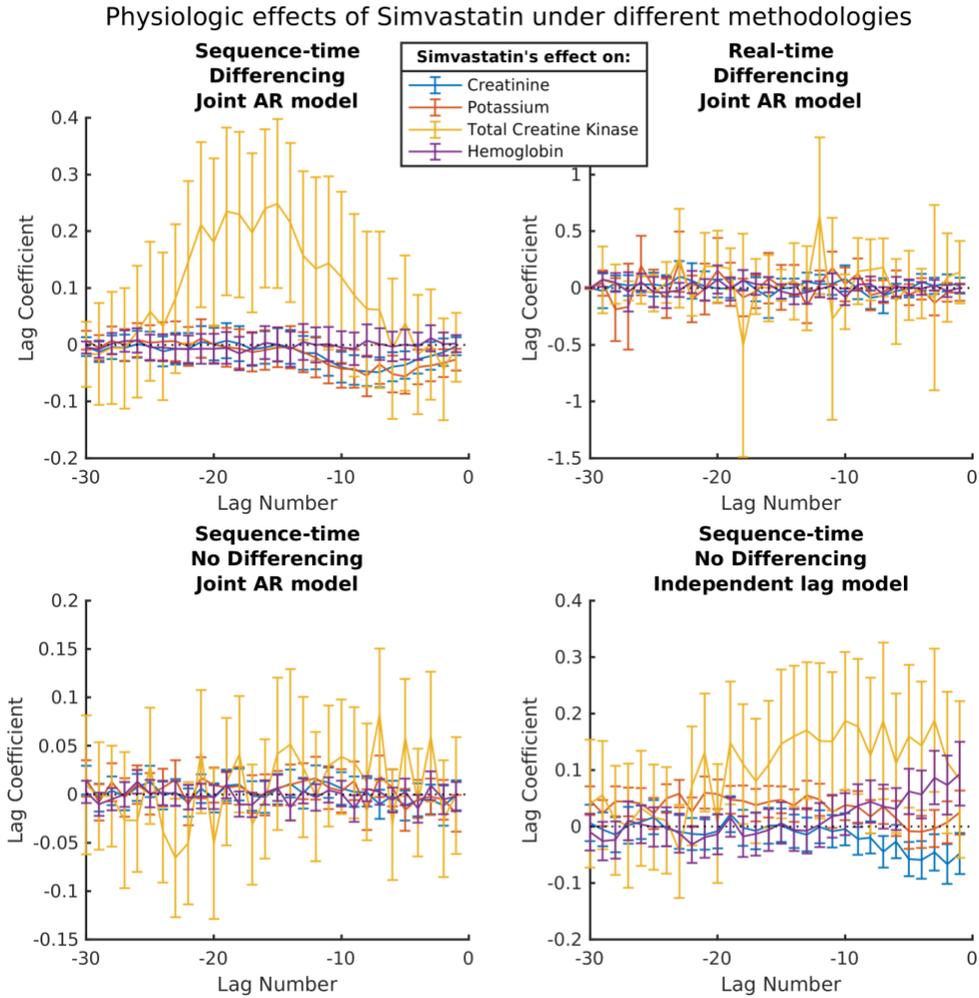

Lagged regression results from four different methodological variations are shown here for simvastatin's effect on creatinine, potassium, total creatine kinase, and hemoglobin. Specifically, we show 4 illustrative combinations of temporal parameterization, differencing, and model choices; in all figures, intra-patient normalization was used, no windowing or binning was performed, and no additional context variable was used.

**Supplementary Figure 3**

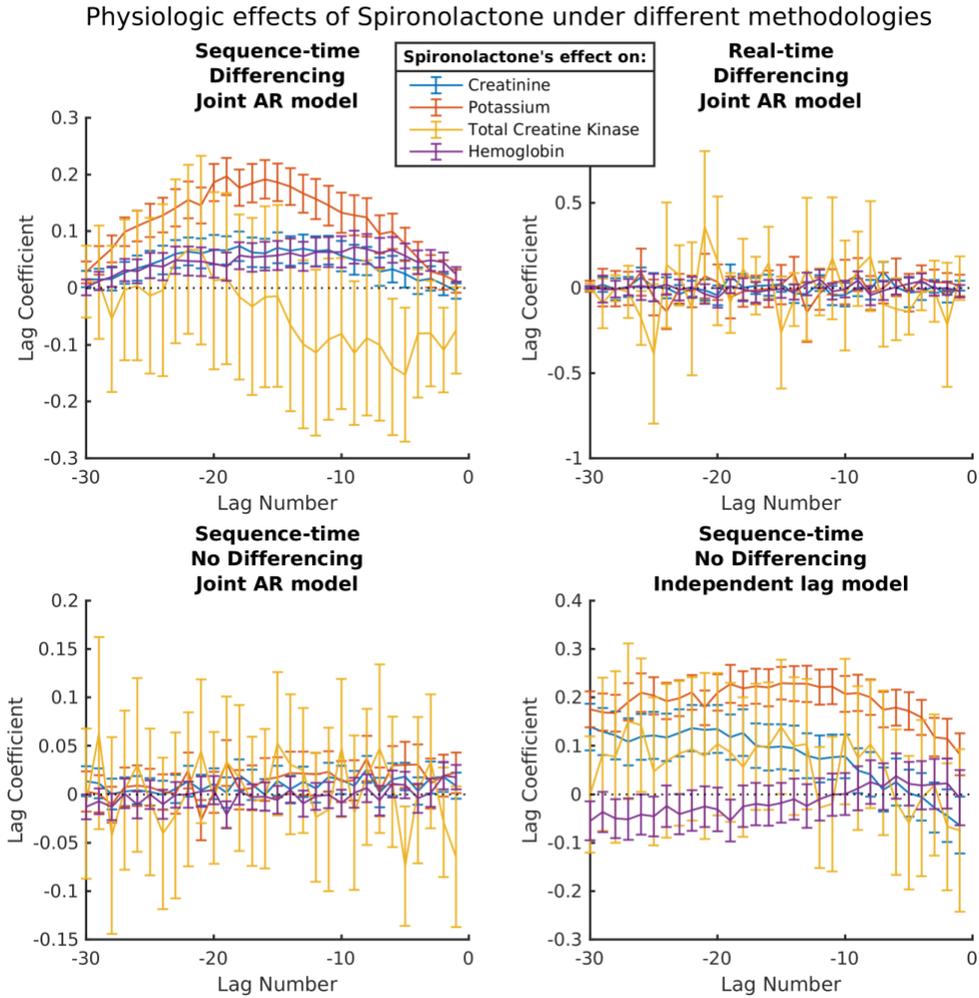

Lagged regression results from four different methodological variations are shown here for spironolactone's effect on creatinine, potassium, total creatine kinase, and hemoglobin. Specifically, we show 4 illustrative combinations of temporal parameterization, differencing, and model choices; in all figures, intra-patient normalization was used, no windowing or binning was performed, and no additional context variable was used.

**Supplementary Figure 4**

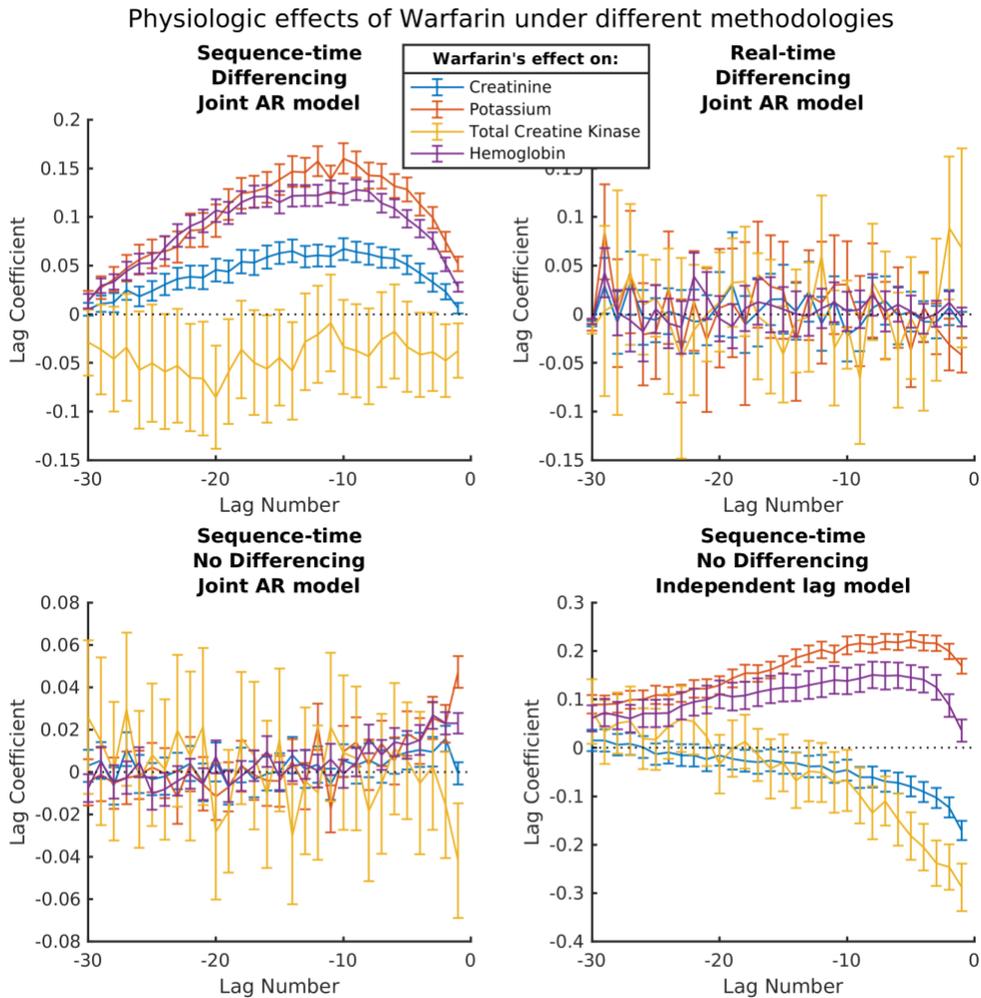

Lagged regression results from four different methodological variations are shown here for warfarin's effect on creatinine, potassium, total creatine kinase, and hemoglobin. Specifically, we show 4 illustrative combinations of temporal parameterization, differencing, and model choices; in all figures, intra-patient normalization was used, no windowing or binning was performed, and no additional context variable was used.

**Supplementary Figure 5**

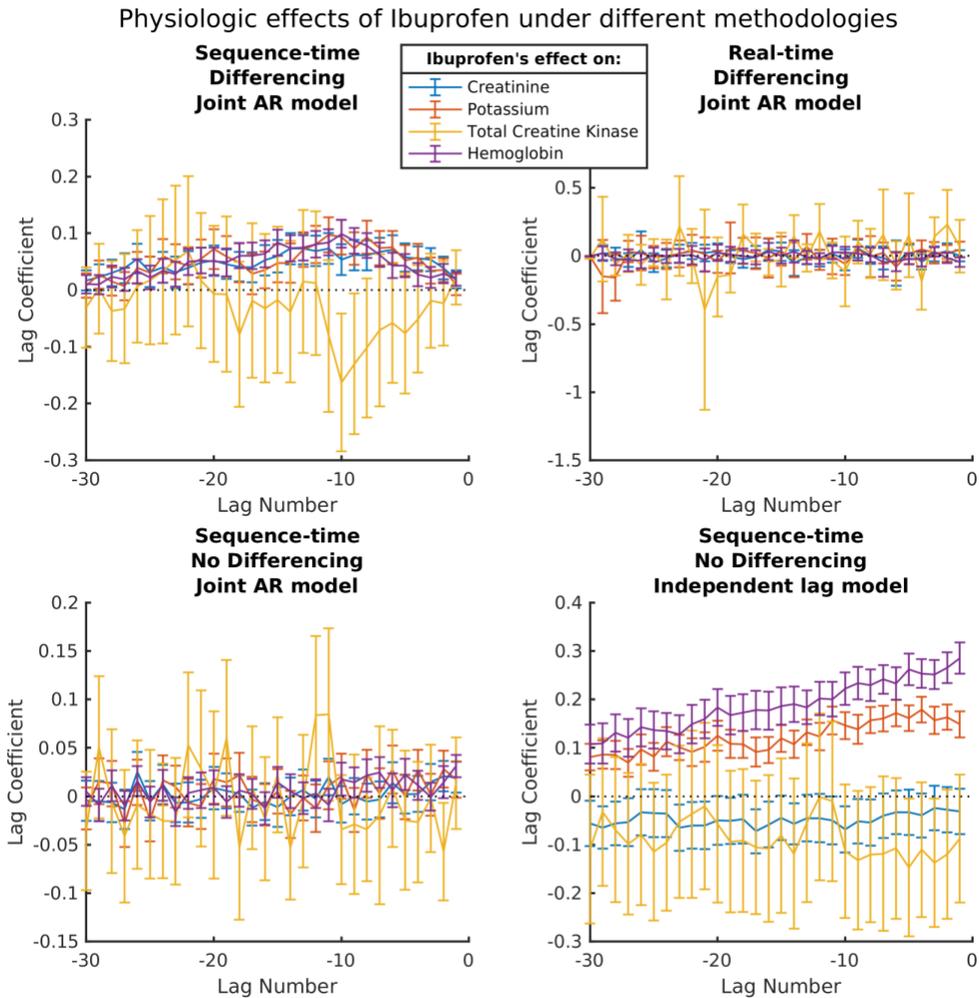

Lagged regression results from four different methodological variations are shown here for ibuprofen's effect on creatinine, potassium, total creatine kinase, and hemoglobin. Specifically, we show 4 illustrative combinations of temporal parameterization, differencing, and model choices; in all figures, intra-patient normalization was used, no windowing or binning was performed, and no additional context variable was used.

**Supplementary Figure 6**

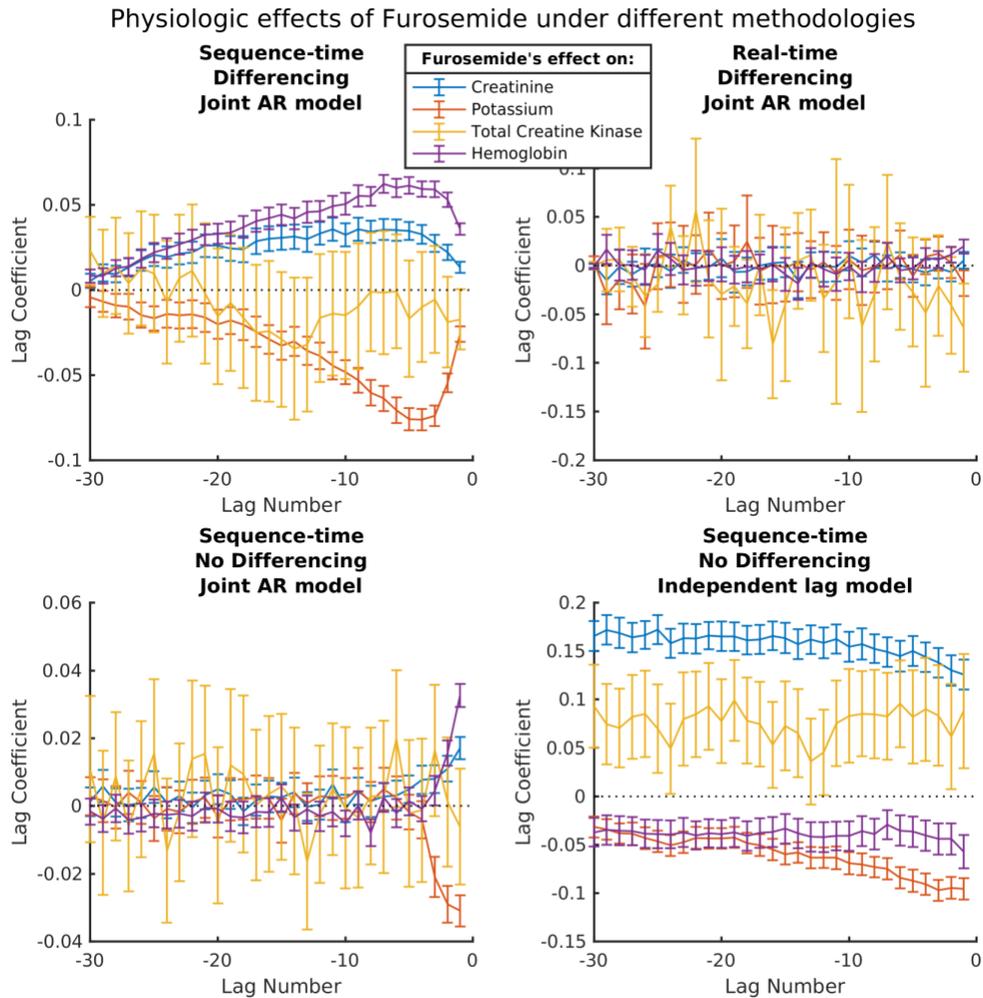

Lagged regression results from four different methodological variations are shown here for furosemide's effect on creatinine, potassium, total creatine kinase, and hemoglobin. Specifically, we show 4 illustrative combinations of temporal parameterization, differencing, and model choices; in all figures, intra-patient normalization was used, no windowing or binning was performed, and no additional context variable was used.

**Supplementary Figure 7**

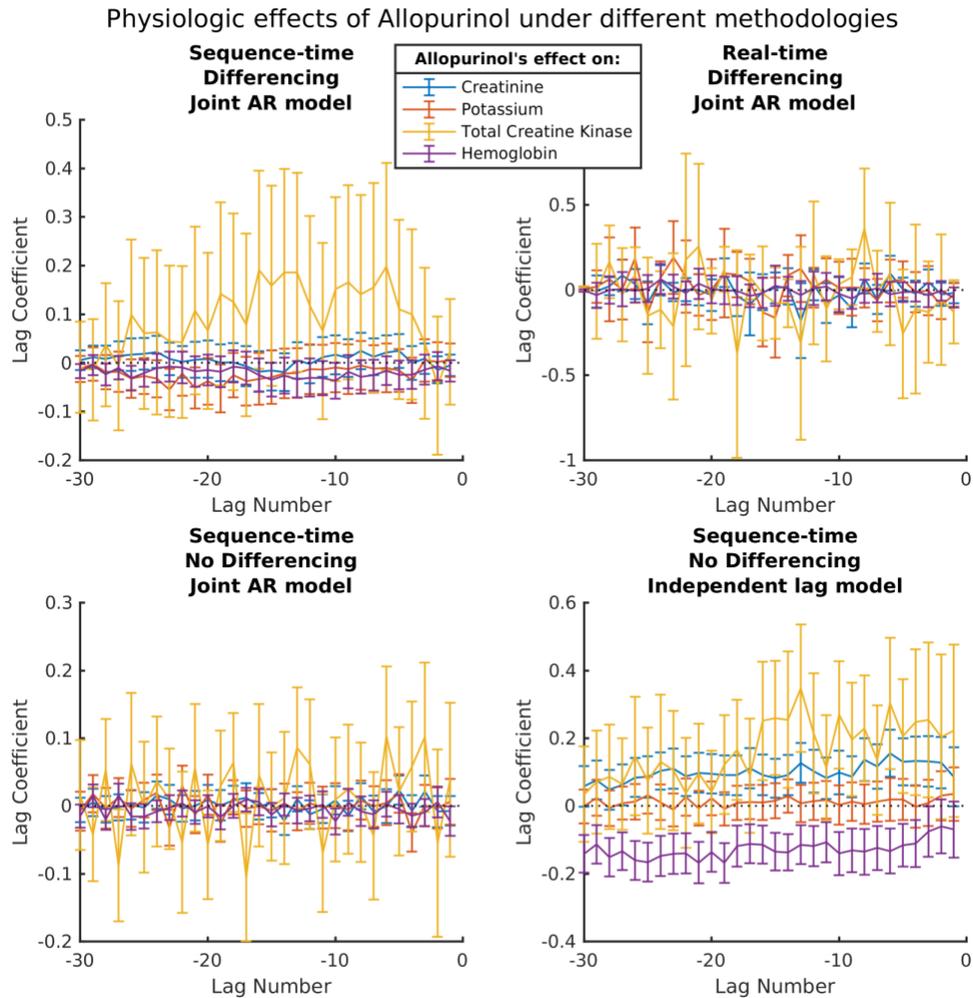

Lagged regression results from four different methodological variations are shown here for allopurinol's effect on creatinine, potassium, total creatine kinase, and hemoglobin. Specifically, we show 4 illustrative combinations of temporal parameterization, differencing, and model choices; in all figures, intra-patient normalization was used, no windowing or binning was performed, and no additional context variable was used.